\documentclass[letter]{aa} 

\usepackage{graphicx}
\usepackage{txfonts}
\usepackage{CJKutf8}
\usepackage{placeins}
\usepackage{lscape}
\begin{document}

   \title{On the N/O abundance ratio and the progenitor mass\\ 
   for the most luminous planetary nebulae of M\,31}

   \author{Toshiya Ueta (\begin{CJK}{UTF8}{min}植田 稔也\end{CJK})\inst{1}
          \and
          Masaaki Otsuka (\begin{CJK}{UTF8}{min}大塚 雅昭\end{CJK})\inst{2}
          }

   \institute{Department of Physics and Astronomy,
              University of Denver,
              2112 E Wesley Ave.,
              Denver, CO 80208, USA\\
              \email{toshiya.ueta@du.edu}
         \and
             Okayama Observatory, 
             Kyoto University, 
             Honjo, Kamogata, Asakuchi, Okayama, 719-0232, Japan\\
             \email{otsuka@kusastro.kyoto-u.ac.jp}
             }

    \titlerunning{}
    \authorrunning{Ueta \& Otsuka}

   \date{Received August 12, 2022; Accepted October 9, 2022}

 
  \abstract
   {Plasma diagnostics are the bases of investigation into the physical and chemical properties 
   of line-emitting gaseous systems.}
   {To perform plasma diagnostics properly, it is essential to correct the input spectrum for extinction properly.
   This is simply because determining 
   the degree of extinction is dependent on the physical properties of the line-emitting gas. 
   Hence, both extinction correction and plasma diagnostics have to be performed simultaneously and self-consistently.}
   {By comparing the results of analyses performed for a sample of nine bright planetary nebulae in M\,31
   with and without the fully iterative self-consistent simultaneous extinction correction and plasma diagnostics,
   we demonstrate how a seemingly benign initial assumptions for the physical conditions of the line-emitting gas 
   in extinction correction would compromise the results of the entire analyses in terms of the extinction,
   electron density/temperature, and ionic/elemental abundances.}
   {While the electron density/temperature are relatively immune to the imposed inconsistent assumptions,
   the compromised extinction would cause systematic offsets in the extinction-corrected line strengths/spectrum, 
   which consequently would impose adverse effects on the resulting ionic and elemental abundances, and 
   other inferences made from the incorrect results.}
   {We find that this M\,31 PN sample simply represents those around the high-mass end of the mass range 
   for low-mass planetary nebula progenitor stars as expected from the existing theoretical models. 
   It appears that the suspicion raised in the previous study -- these PNe 
   being anomalously nitrogen overabundant for the expected progenitor mass range -- 
   is simply caused by the apparent underestimate in extinction that originates 
   from the imposed inconsistent assumptions in extinction correction.
   In a larger context, the results of plasma diagnostics in the literature 
   without seeking simultaneous self-consistency
   with extinction correction have to be handled cautiously.
   Ideally, such previous results should be re-evaluated by seeking simultaneous self-consistency.}
   

   \maketitle

\section{Introduction}

In a recent article ``On the most luminous planetary nebulae of M\,31,'' 
\citet[][hereafter GR22]{gr2022} have presented the results of plasma diagnostics 
of planetary nebulae (PNe) in the Andromeda Galaxy (M\,31).
The analyses were based on optical spectra of nine bright PNe 
(the brightest four, plus five others as control cases) taken with the OSIRIS instrument 
at the 10.4 m Gran Telescopio Canarias (GTC) supplemented by archival spectra.
Their study was aimed at investigating the physico-chemical properties of
these PNe and the progenitor mass of the central stars expected to be 
at the tip of the PN luminosity function (PNLF; \citealt{jacoby1980}).

The nine bright PNe in M\,31 in question have been found to have achieved 
the maximum temperature as suggested by the post-AGB evolutionary tracks for stars 
with initial mass of $\sim$1.5\,M$_{\odot}$, 
providing spectroscopic constraints for the stellar progenitors that define the PNLF cutoff 
for M\,31 and for other similar star-forming galaxies. 
However, these PNe have also been found with the N/O abundance ratio that is 1.5 to 3 times larger 
than predicted for such stars (of $\sim$1.5\,M$_{\odot}$ initial mass), 
indicating possible limitations for the existing theoretical models.

Meanwhile, our recent examination on a specific PN case has indicated that,
unless both extinction correction and plasma diagnostics are performed simultaneously and self-consistently,
the results of the subsequent abundance analyses can be off by tens of percents \citep{ueta2021}. 
For emission-line objects, the degree of extinction, 
$c({\rm H}\beta)$,\footnote{The base-10 power-law index $c(\lambda)$ as in $I(\lambda) = I_{0}(\lambda) \times 10^{-c(\lambda)} =  I_{0}(\lambda) \times 10^{-c({\rm H}\beta)(1+f(\lambda))}$ referenced at H$\beta$.
Here, $I(\lambda)$ and $I_0(\lambda)$ are the observed (attenuated) and intrinsic (unattenuated) specific flux at $\lambda$
and $f(\lambda)$ refers to the adopted extinction law.}
is typically estimated by comparing the observed (attenuated) 
diagnostic line flux ratio (usually of a Balmer H line pair, most often the H$\alpha$-to-H$\beta$ line flux ratio) 
with its theoretical (unattenuated) counterpart. 

This is a classic chicken-and-egg problem, because the theoretical H line ratio
-- the basis for extinction correction -- is actually dependent on 
the electron density ($n_{\rm e}$) and  temperature ($T_{\rm e}$) of the line emitting gas,  
which are the very quantities to be determined via plasma diagnostics 
that must be performed {\it after extinction is properly corrected}. 
Therefore, to achieve the maximum self-consistency, 
the theoretical H line ratio used in extinction correction must be updated 
simultaneously and iteratively with $n_{\rm e}$ and $T_{\rm e}$ 
obtained in plasma diagnostics until they converge to the optimum values.
However, such an approach 
has been seldom practiced in the literature for some unknown reasons, 
most likely to alleviate the volume of non-linear numerical calculations 
when computational resources were still scarce \citep{ueta2022}.
At any rate, it is simply incorrect to perform extinction correction just once
with some assumed $n_{\rm e}$ and $T_{\rm e}$ before performing plasma diagnostics.

In the recent M\,31 work by GR22, the H$\alpha$-to-H$\beta$ line ratio was fixed to 2.86 
(equivalent to assuming $T_{\rm e}=10^4$\,K and $n_{\rm e} = 10^3$\,cm$^{-3}$) 
in deriving the extinction, $c({\rm H}\beta)$, irrespective of what the subsequent plasma diagnostics 
suggested for $n_{\rm e}$ and $T_{\rm e}$. 
Here, we opt to demonstrate
how this widely adopted initial assumption 
-- the fixed H$\alpha$-to-H$\beta$ line ratio -- 
would influence the results of extinction correction and plasma diagnostics 
as well as any subsequent inferences based on the results of these analyses.

\section{Analyses}

We adopt the measured (uncorrected) line fluxes of these nine PNe in M\,31 
as presented by GR22 (from their Table\,A.1).
Regretfully, these fluxes are given in only three significant figures and 
without measurement uncertainties. 
As for uncertainties, GR22 stated that ``total errors were determined from the quadratic propagation 
of the measured statistical errors of the spectra, which in general decrease from blue toward red, 
plus an estimated additional 3\,\% of the measured flux in order to account for systematic errors, 
which include continuum determination, flux calibration and, although less important, wavelength 
calibration uncertainties.'' 
Thus, without exact uncertainties to quote, 
we choose to assume 5\,\% uncertainty across the spectrum.

For the rest of the analyses, 
we follow the same procedure as outlined by GR22 as long as they are described.
The apparent deviation is adopting the self-consistent H$\alpha$-to-H$\beta$ line ratio 
in extinction correction according to the updated $n_{\rm e}$ and $T_{\rm e}$ from plasma diagnostics 
in order to guarantee simultaneous self-consistency between extinction correction and plasma diagnostics.
Both extinction correction and plasma diagnostics are performed with {\sc PyNeb} \citep{pyneb} 
using the same atomic parameter set (PYNEB\_18\_01, as summarized in Tables 4 and 6 by GR22). 
Uncertainties are assessed statistically by computing 1500 Monte Carlo simulations 
while allowing input line fluxes with Gaussian uncertainties.

Also, we assume a two-component nebula with high- and low-excitation regions 
characterized by the cut-off ionization potential (IP) of 17\,eV. 
Transitions above this IP cut-off are considered of high-excitation and computed $n_{\rm e}$ and $T_{\rm e}$ 
derived from the [O\,{\sc iii}] $\lambda$4363/$\lambda$4959 and [Ar\,{\sc iv}] $\lambda$4711/$\lambda$4740 
diagnostic line ratios, 
while those below (of low-excitation) are calculated with 
the [N\,{\sc ii}] $\lambda$5755/$\lambda$6548 and [S\,{\sc ii}] $\lambda$6716/$\lambda$6731 diagnostic line ratios. 
A minor exception adopted by GR22 was that they used $T_{\rm e}$([O\,{\sc iii}]) even for low-excitation regions 
when $T{\rm e}$([N\,{\sc ii}]) resulted with uncertainties greater than $\sim$2,000\,K (for M2068 and M1675). 
As shown below, we do not make any such exceptions 
because uncertainties are below 1,200\,K in our analyses.

\section{Extinction Correction}

To verify that the results of our analyses can be compared squarely with those by GR22, 
we first emulate only a single run of extinction correction and plasma diagnostics 
as was performed by GR22.
This means that $T_{\rm e}$ and $n_{\rm e}$ are fixed in extinction correction
at $10^4$\,K and $10^3$\,cm$^{-3}$, respectively 
(equivalent to adopting the fixed theoretical H$\alpha$-to-H$\beta$ line ratio of 2.86)
under the same extinction law \citep{ccm1989} and 
the total-to-selective extinction ratio, $R_{\rm V}$, of 3.1.
Then, $T_{\rm e}$ and $n_{\rm e}$ are iterated only in plasma diagnostics.
In other words, 
plasma diagnostic would find optimum $T_{\rm e}$ and $n_{\rm e}$ via iteration,
but only after diagnostic lines are extinction-corrected with $c({\rm H}\beta)$
based on the pre-fixed $T_{\rm e} = 10^4$\,K and $n_{\rm e} = 10^3$\,cm$^{-3}$
(hence, the optimum $T_{\rm e}$ and $n_{\rm e}$ are most likely 
inconsistent with the presumed $T_{\rm e}$ and $n_{\rm e}$).
With this emulated procedure established,
the only difference is the adopted input flux uncertainties. 
As shown in Table\,\ref{tab:cHb},
our $c({\rm H}\beta)$ values (``Emulated'') in this verification are generally consistent 
with those obtained by GR22 (within $94.0\pm6.5$\,\%). 
Thus, we consider that our calculations do reproduce the results obtained by GR22 reasonably well, 
and hence, all comparisons that follow are indeed valid. 

\begin{table}
	\centering
	\caption{The extinction $c({\rm H}\beta)$ toward nine bright PNe in M\,31. 
	Listed are values as computed by GR22, GR22-emulated by us (Emulated),
	and fully iterated by us (Full).
	Also listed are the GDRE values.}
	\label{tab:cHb}
	\begin{tabular}{lcccc} 
		\hline\hline
		PN & GR22 & Emulated & Full & GDRE \\
		\hline
		M1687 & 0.20$\pm$0.06 & 0.20$\pm$0.03 & 0.24$\pm$0.02 & 0.26$\pm$0.06 \\
		M2068 & 0.13$\pm$0.06 & 0.15$\pm$0.02 & 0.37$\pm$0.02 & 0.25$\pm$0.07 \\
		M2538 & 0.00$\pm$0.00 & 0.06$\pm$0.01 & 0.35$\pm$0.10 & 0.04$\pm$0.01 \\
		M50   & 0.19$\pm$0.06 & 0.20$\pm$0.03 & 0.28$\pm$0.02 & 0.11$\pm$0.01 \\
		M1596 & 0.17$\pm$0.06 & 0.18$\pm$0.02 & 0.20$\pm$0.02 & 0.10$\pm$0.03 \\
		M2471 & 0.04$\pm$0.06 & 0.09$\pm$0.01 & 0.34$\pm$0.04 & 0.07$\pm$0.01 \\
		M2860 & 0.10$\pm$0.06 & 0.12$\pm$0.01 & 0.54$\pm$0.08 & 0.08$\pm$0.01 \\
		M1074 & 0.11$\pm$0.06 & 0.12$\pm$0.01 & 0.21$\pm$0.02 & 0.31$\pm$0.06 \\
		M1675 & 0.23$\pm$0.06 & 0.23$\pm$0.03 & 0.54$\pm$0.05 & 0.26$\pm$0.06 \\
		\hline
	\end{tabular}
\end{table}

In Table\,\ref{tab:cHb},
we also quote the $c({\rm H}\beta)$ values according to the NASA/IPAC Infrared 
Science Archive Galactic Dust Reddening and Extinction (GDRE) 
service (based on the SDSS spectra; \citealt{sf2011}).\footnote{https://irsa.ipac.caltech.edu/applications/DUST/}
The GDRE values are the mean $c({\rm H}\beta)$ value within 
a sampling circle of 5\,arcmin radius
toward each of the nine M\,31 PNe, accounting for extinction 
from all but the circumstellar component for each PN, i.e.,
the interstellar component in both the Milky Way and M\,31 and
the intragalactic component between the Milky Way and M\,31. 
Thus, the GDRE value may overestimate the M\,31 interstellar component
because it accounts for extinction along the line of sight {\it even beyond} the target PN.
At any rate, our fully iterated $c({\rm H}\beta)$ values (``Full'') turn out 
generally greater than the GDRE values, 
corroborating the presence of non-zero circumstellar extinction component
that varies a lot from object to object (from 0.46 to 0.1, corresponding to 65 to 20\,\% attenuation).
This means that individual circumstellar $c({\rm H}\beta)$ value cannot just be neglected
or generically assumed: it must be computed for each object self-consistently.
At a minimum, extinction correction with fixed $n_{\rm e}$ and $T_{\rm e}$ 
appears already dubious.

On the whole, direct comparisons between the GR22 values and our fully iterative results
show that $c({\rm H}\beta)$ by GR22 was more than 50\,\% underestimated ($52.5\pm31.2$\,\%) 
with individual variations from 12.6 to 88.3\,\%. 
Because $c(\lambda)$ is the base-10 power-law index varying non-linearly across the spectrum,
the impact of offsets in $c(\lambda)$ is hard to gauge unless actually calculated.
The extinction-corrected line fluxes based on the fully iterative $c({\rm H}\beta)$ are listed in 
Tables\,\ref{tab:flx1} and \ref{tab:flx2},
along with the original-to-revised ratio for comparison.
We see that 
the GR22 fluxes were underestimated (not sufficiently extinction-corrected) 
by up to 20\,\% at the shortest [O\,{\sc ii}] $\lambda$3727 line
and overestimated (excessively extinction-corrected) by up to 59\,\% at the longest [Ar\,{\sc iii}] $\lambda$7751 line.

\section{Plasma Diagnostics: Physical Conditions}

The resulting optimized $n_{\rm e}$ and $T_{\rm e}$,
after full iteration of extinction correction and plasma diagnostics,
are summarized in Table\,\ref{tab:nT}. 
Convergence is achieved in less than or equal to 4 iterations in all cases.
The ratio column shows comparisons between the GR22 values to ours. 
No PN shows more than two-$\sigma$ deviations after one full iteration 
in both $T_{\rm e}$([O\,{\sc iii}]) and $T_{\rm e}$([N\,{\sc ii}]). 
This stability of $T_{\rm e}$ seems to stem from the fact that 
the $T_{\rm e}$-diagnostic is not only insensitive to $n_{\rm e}$ 
(which is why it works as the $T_{\rm e}$ diagnostic), but also 
varies only slightly in $T_{\rm e}$  
in the $n_{\rm e}$--$T_{\rm e}$ diagnostic plane 
for the given diagnostic line ratio. 

The $n_{\rm e}$([S\,{\sc ii}]) values agree remarkably well ($100\pm3$\,\% agreement), 
neglecting two very deviant cases 
(substantially overestimated by 177 and 235,\% for M2068 and M1675, respectively). 
For these two deviant PNe, GR22 found very large uncertainties (not explicitly defined).
So do we with about 50\,\% uncertainties. 
Meanwhile, the $n_{\rm e}$([Ar\,{\sc iv}]) values were overestimated by GR22 ($141.2\pm24.8$\,\%). 
This seems to have been caused by the $c({\rm H}\beta)$ underestimates 
in mitigating He\,{\sc i} $\lambda$4713 line contamination in the [Ar\,{\sc iv}] $\lambda$4711 line flux,
as will be described below. 

Given the overall rough consistency among the $n_{\rm e}$ and $T_{\rm e}$ values 
obtained with and without fully iterative calculations,
especially the rather invariant $T_{\rm e}$ results,
one may be tempted to forgo seeking convergence in extinction correction.
In determining $n_{\rm e}$ and $T_{\rm e}$ in plasma diagnostics, 
one tends to think of it as ``fitting'' for which the optimum values are found 
from a free excursion in the $n_{\rm e}$-$T_{\rm e}$ space. 
In reality, however, the applicable range of $n_{\rm e}$ is also rather restricted,
because the $n_{\rm e}$-diagnostic curve varies from one limiting ratio to another within a narrow range of $n_{\rm e}$
given the choice of the $n_{\rm e}$-diagnostic line.

In the $n_{\rm e}$-$T_{\rm e}$ space, 
{\it given the measured line ratios of the adopted diagnostic lines},
where the solution exists (i.e.\ where diagnostic curves intersect) is already set.
Hence, what ``fitting'' does in the $n_{\rm e}$-$T_{\rm e}$ space is to find the optimally closest ($n_{\rm e}$, $T_{\rm e}$) point 
to the intersection of the diagnostic curves within measurement uncertainties.
In this sense, it is expected that plasma diagnostics with the measured line ratios
would yield more or less the same $n_{\rm e}$ and $T_{\rm e}$ values,
irrespective of the rigorousness in the preceding extinction correction. 
A true exploration of the $n_{\rm e}$-$T_{\rm e}$ space can only be done 
by allowing the measured diagnostic line ratios vary according to $c({\rm H}\beta)$.
This relative insensitivity of $n_{\rm e}$ and $T_{\rm e}$ against plasma diagnostics 
probably fostered a sense of ``lessez-faire'' in the community to the extent that 
the theoretical H$\alpha$-to-H$\beta$ line ratio of 2.86 is referred to as ``canonical''
when there is no canonicity whatsoever to this value. 
As we will see below, omitting iteration in extinction correction is not at all a viable option,
because the resulting $c({\rm H}\beta)$ offset would seriously impact the rest of the plasma diagnostics. 

\section{Plasma Diagnostics: Ionic Abundances}

As for calculating ionic abundances with {\sc PyNeb}, 
we again follow GR22 as much as possible even though sometimes the procedure was not explicitly described.
The derived ionic abundances for all the input collisionally excited lines and He recombination lines are summarized in Tables\,\ref{tab:ion1} and \ref{tab:ion2}, 
even though GR22 presented only a subset of lines in their Table\,A.2. 

Theoretically speaking, 
ionic abundances derived from the same ionic transitions should turn out identical given the adopted $n_{\rm e}$ and $T_{\rm e}$. 
In the literature, this is not necessarily accomplished.
Discrepancies among derived ionic abundances for a particular ionic species are
often attributed to local temperature (and density) fluctuations in the line-emitting gas.
This attribution is actually very odd, 
especially when the fixed ``canonical'' H$\alpha$-to-H$\beta$ line ratio of 2.86 is adopted
(i.e., the uniform $n_{\rm e}$ at $10^3$\,cm$^{-3}$ and $T_{\rm e}$ at $10^4$\,K are imposed).
If we are to believe varying $n_{\rm e}$ and $T_{\rm e}$ in the target nebula along the line of sight,
assuming the ``canonical'' uniformity in $n_{\rm e}$ and $T_{\rm e}$ in extinction correction is 
equivalent to {\it injecting inconsistency} into the analyses in the first place.
This is even worse in analysing 2-D plasma diagnostics with line emission maps,
in which spatial variation is surely expected by default.
At any rate, when ionic abundances derived from multiple lines of an ionic specie vary,
some sort of averaging needs to be done to define a representative value.
GR22 took the straight average, while we take the uncertainty-weighted mean.

Our derived ionic abundances show good self-consistency nearly across the board: 
the derived abundances for most of the ionic species all agree within uncertainties. 
In all major transitions to be used in calculating the total elemental abundances 
(He$^{+}$, He$^{2+}$, O$^{+}$, and O$^{2+}$) 
with the help of the ionization correction factors (ICFs; \citealt{di2014}), 
the derived ionic abundances are consistent with each other except for O$^{+}$ from M1675, 
for which the [O\,{\sc ii}] $\lambda$7320 line was not measured 
among the trio of [O\,{\sc ii}] lines at 3727, 7320, and 7330\,{\AA}. 
Among this O$^{+}$ trio, the [O\,{\sc ii}] $\lambda$3727 
line\footnote{The [O\,{\sc ii}] $\lambda$3727 line is actually a blend of two lines at 3726 and 3729\,{\AA}. 
In our calculations, it is treated as a blend.} 
is located at the blue-end of the detector bandwidth,
while the [O\,{\sc ii}] $\lambda$7330 is at the red-end.
Not knowing which is more accurate between the two, 
we simply take the uncertainty-weighted mean as others.

GR22 obtained O$^{+}$ abundances that were very discrepant (see their Table A.2).
The O$^{+}$ $\lambda$3772 abundance came out to be less than half of the O$^{+}$ $\lambda$7320/7330 abundances,
except for the recurring anomalous case of M2068 (for which the former was 2.5 times greater).
There is a simple explanation as to why the O$^{+}$ $\lambda$3772 abundance came out much smaller:
the underestimated $c({\rm H}\beta)$.
Here, it is reminded that what counts is a ratio of given line flux {\it relative} to H$\beta$ flux.
The [O\,{\sc ii}] $\lambda$3772 is on the short-ward of H$\beta$, and hence,
the extinction-corrected flux is underestimated 
(i.e., not corrected enough by the underestimated $c({\rm H}\beta)$).
On the other hand, the [O\,{\sc ii}] $\lambda$7320/7330 are on the long-ward of H$\beta$, and hence, 
the extinction-corrected flux is overestimated
(i.e., not reduced enough by the underestimated $c({\rm H}\beta)$).
As a result, the underestimated $c({\rm H}\beta)$ leads to 
under/over-estimated abundances on the opposite sides of the reference H$\beta$ wavelength 
(as seen in Tables\,\ref{tab:ion1} and \ref{tab:ion2}).
The O$^{+}$ abundance adopted by GR22 appears to have been biased toward an overestimate,
as the [O\,{\sc ii}] $\lambda$7320/7330 lines are much farther away from H$\beta$
than the [O\,{\sc ii}] $\lambda$3772 line.

The He$^{+}$ and He$^{2+}$ abundances by GR22 suffered from the same issue,
because the He$^{+}$ abundance was adopted from the He\,{\sc i} $\lambda$5876
(hence, tended to be overestimated)  
and the He$^{2+}$ abundance from the He\,{\sc i} $\lambda$4686
(hence, tended to be underestimated). 
This trending is also seen in Tables\,\ref{tab:ion1} and \ref{tab:ion2}.
These observations reveal that anomalies in abundance analyses would arise not from 
offsets in the $n_{\rm e}$ and $T_{\rm e}$ values, but rather from the possible under/over-estimate
by the underestimated $c({\rm H}\beta)$ on either side of the reference wavelength of H$\beta$ at 4861\,{\AA}.
GR22 also used the corrected He$^{+}$ and He$^{2+}$ fluxes to constrain $T_{\rm eff}$ 
in comparison with theoretical models.
Hence, the impact of an $c({\rm H}\beta)$ offset can be seen at many different places in the analyses.

As for the rest of the observed lines, we find derived abundances somewhat discrepant for 
Ar$^{2+}$ (7136/7751\,{\AA}), 
Cl$^{2+}$ (5518/5538\,{\AA}), and 
S$^{+}$ (4069/4076\,{\AA}). 
The [Ar\,{\sc iii}] $\lambda$7751 line is located at the red-end of the bandwidth 
and appears relatively strongly affected by the atmospheric absorption 
(by the O$_{2}$ band around 8000\,{\AA}, verified in the OSIRIS 2-D spectrum itself). 
The [Cl\,{\sc iii}] lines are intrinsically weak (about $\sim$1 when $I({\rm H}\beta) = 100$). 
As for the [S\,{\sc ii}] lines, GR22 referred the [S\,{\sc ii}] $\lambda$4076 line as the [S\,{\sc ii}] $\lambda$4071 line, 
and hence, the line identification might have been compromised.
As these ionic abundances are not directly involved in the subsequent elemental abundance calculations, 
we leave them as they are.
However, these ionic abundances, if incorrect, will affect the corresponding elemental abundances in the end.
At any event, there always seem some reasonable explanations 
as to why the derived abundances are discrepant.

However, there is more subtle but involved complications in the [Ar\,{\sc iv}] $\lambda$4711/4740 lines,
which play an important role in the present analyses as the high-excitation $n_{\rm e}$-diagnostic lines.
The [Ar\,{\sc iv}] $\lambda$4711 is contaminated by the neighboring He\,{\sc i} $\lambda$4713 line,
whose strength needs to be estimated by scaling the He\,{\sc i} $\lambda$5876 measurement
(the strongest, hence, the most reliable of all the detected He lines)
at the corresponding $n_{\rm e}$ and $T_{\rm e}$.
Thus, the [Ar\,{\sc iv}] $\lambda$4711 measurement is unavoidably underestimated 
when the He\,{\sc i} $\lambda$5876 measurement is overestimated by the underestimated $c({\rm H}\beta)$.
Hence, the [Ar\,{\sc iv}] $\lambda$4711 line flux corrected for the He\,{\sc i} $\lambda$4713 contamination
by GR22 was most likely doubly affected by this mechanism.
This indeed artificially reduced the [Ar\,{\sc iv}] $\lambda$4711/$\lambda$4740 ratio for GR22,
forcing the resulting $n_{\rm e}$ become larger by $141.2\pm24.8$\,\%.

By the same token,
the [O\,{\sc iii}] $\lambda$4363/$\lambda$4959 $T_{\rm e}$-diagnostic ratio obtained by GR22
was at least mildly affected by the underestimated $c({\rm H}\beta)$:
the ratio was artificially reduced, forcing the resulting $T_{\rm e}$ become smaller
by $96.5\pm2.0$\,\% (Table\,\ref{tab:nT}),
which is still an underestimate.
Conversely, other $n_{\rm e}$ and $T_{\rm e}$ diagnostics did not seem to suffer from the underestimated $c({\rm H}\beta)$,
because in these cases
with the [N\,{\sc ii}] $\lambda$5755/$\lambda$6548
and [S\,{\sc ii}] $\lambda$6716/$\lambda$6731 lines
for the low-excitation region
the lines are on the same side of H$\beta$ and not far from each other.
Thus, the effects of the underestimated $c({\rm H}\beta)$ were marginalised.

\section{Plasma Diagnostics: Elemental Abundances}

As the final step of plasma diagnostics, 
the total elemental abundances are computed from the derived ionic abundances
in terms of $A({\rm X}) = 12+\log({\rm X}/{\rm H})$ (Table\,\ref{tab:abun}).
First, we emulate GR22 by adopting the same ICFs \citep{di2014}
with the derived ionic abundances of 
O$^{+}$ and O$^{2+}$, N$^{+}$, Ne$^{2+}$, S$^{+}$, Ar$^{2+}$, and Cl$^{2+}$.
However, there are unused ionic species such as S$^{2+}$, Ar$^{3+}$, and Ar$^{4+}$ in the analyses.
In fact, it is rather strange that the Ar$^{3+}$ abundance was not used by GR22,
because [Ar\,{\sc IV}] was adopted as the $n_{\rm e}$ diagnostic.
Because the Ar abundance can work as an important metallicity indicator,
there is no reason not to adopt all three measured Ar ionic species 
to reduce reliance/uncertainty of the ICF.
Hence, we also compute the total elemental abundances
by adopting all observed ionic abundances.
In this case, the adopted ICFs are still based on those suggested by \citet{di2014} 
except for Ar, for which we assume ${\rm Ar} = {\rm Ar}^{2+} + {\rm Ar}^{3+} + {\rm Ar}^{4+}$. 

In the ICF formulation,
to yield the total elemental abundance,
the sum of the observed ionic abundances for a specific element 
has to be scaled to account for the unobserved ionic species.
These scaling factors are empirically defined as functions of He and O 
ionic abundances (He$^{+}$, He$^{2+}$, O$^{+}$, and O$^{2+}$; \citealt{di2014}).
This means that the reliability of the observed ionic abundances 
depends on the derived abundances of these He and O ionic species.

The O elemental abundance is based on the observed O$^{+}$ and O$^{2+}$ ionic abundances.
Both of the O$^{+}$ and O$^{2+}$ abundances are based on three lines,
[O\,{\sc ii}] lines at 3727, 7320, and 7330\,{\AA} and
[O\,{\sc iii}] lines at 4363, 4959, and 5007\,{\AA}, respectively.
As these lines are distributed on both sides of H$\beta$ at 4861\,{\AA},
the effects of the underestimated $c({\rm H}\beta)$ was most likely marginal 
even in the results by GR22.
Indeed, the median O abundance among the nine PN sample obtained by GR22 was $8.63 \pm 0.09$,
while ours is $8.59 \pm 0.08$: fairly consistent with relatively small uncertainties.
This does not necessarily mean that the O abundance is insensitive to the $c({\rm H}\beta)$ discrepancy.
The present PN sample is of high-excitation, and hence, their O abundance is 
relatively less dependent on the more uncertain low-excitation O$^{+}$ abundance.
If targets are of low-excitation, the relative importance of the more uncertain O$^{+}$ abundance
would be greater, and hence, the O abundance would have been compromised by the $c({\rm H}\beta)$ discrepancy.

On the other hand, the N elemental abundance is based solely on the observed N$^{+}$ ionic abundance.
All three [N\,{\sc ii}] lines at 5755, 6548, and 6583\,{\AA} are located on the red-side of H$\beta$.
Hence, 
their line fluxes and ionic abundances were most likely overestimated by GR22 
because of the underestimated $c({\rm H}\beta)$.
The median N elemental abundance obtained was $8.27 \pm 0.38$ by GR22,
while ours is $7.99 \pm 0.35$: this is a factor of 1.9 difference.
While the two are statistically indifferent given the relatively large deviation,
the difference in the elemental abundance amounts to nearly 90\,\%.
When both of the N and O elemental abundances are combined to assess the N/O abundance ratio,
the values come out to be $0.40\pm0.30$ (by GR22) and $0.28\pm0.16$ (by us).
Again, these are statistically indifferent, 
but the absolute difference is an overestimate at nearly 30\,\%.

It was further argued by GR22 that
the derived N/O ratio for the M\,31 PN sample 
was 1.5--3 times greater than expected 
for PNe of the $\sim$1.5\,M$_{\odot}$ initial mass
based on comparisons with theoretical models \citep{kl2016,mb2016,vk2018},
even referring to possible limitations of the existing models.
However, 
we can simply interpret this as another consequence of the underestimated $c({\rm H}\beta)$,
which artificially {\it reddened} the whole spectra of target PNe.

\begin{table}
	\centering
	\caption{Comparison of the best-fit 
	luminosity ($L_{\ast}$ in $\log(L_{\ast}/L_{\odot})$), 
	surface temperature ($T_{\rm eff}$ in $\log(T_{\rm eff})$), 
	and initial mass ($M_{i}$ in $M_{\odot}$)
	of the central star for the M\,31 PN sample, 
	based on {\sc CLOUDY} model fitting constrained by the extinction-corrected line fluxes
	by GR22 (the solar metallicity of $Z=0.02$ assumed for all) and us 
	(an appropriate metallicity in the range of $Z = 0.003 - 0.009$ informed from abundance analyses adopted for each PN; 
	Otsuka \& Ueta {\it in prep.}).}
	\label{tab:model}
	\begin{tabular}{lccccccc} 
		\hline\hline
		& \multicolumn{3}{c}{GR22} && \multicolumn{3}{c}{Otsuka \& Ueta}\\
		\cline{2-4}\cline{6-8}
        PN & $L_{\ast}$ & $T_{\rm eff}$ & $M_{i}$ && $L_{\ast}$ & $T_{\rm eff}$ & $M_{i}$ \\
		\hline
		M1687 & 3.66 & 5.06 & 1.48 && 3.87 & 5.01 &  1.9 \\
		M2068 & 3.62 & 5.05 & 1.34 && 4.04 & 5.01 &  2.4 \\
		M2538 & 3.65 & 5.11 & 1.56 && 4.03 & 5.07 &  2.4  \\
		M50   & 3.59 & 5.13 & 1.42 && 3.88 & 5.08 &  2.0  \\
		M1596 & 3.49 & 5.22 & 1.76 && 3.75 & 5.11 &  1.6  \\
		M2471 & 3.50 & 5.20 & 1.70 && 3.82 & 5.10 &  1.7  \\
		M2860 & 3.51 & 5.08 & 1.20 && 3.97 & 5.03 &  2.2 \\
		M1074 & 3.50 & 5.03 & 1.12 && 3.76 & 5.00 &  1.4 \\
		M1675 & 3.60 & 5.10 & 1.39 && 4.05 & 5.05 &  2.6 \\
		\hline
	\end{tabular}
\end{table}

Table\,\ref{tab:model} lists the best-fit luminosity ($L_{\ast}$),  
surface temperature ($T_{\rm eff}$), and 
initial mass ($M_{i}$) of the central star, 
obtained via {\sc CLOUDY} model fitting 
by GR22 (with the $c({\rm H}\beta)$ underestimate, and the fixed solar metallicity of $Z=0.02$ assumed for all PNe) 
and by us (without the $c({\rm H}\beta)$ underestimate, and 
an appropriate metallicity in the range of $Z = 0.003 - 0.009$ informed from abundance analyses adopted for each PN; 
Otsuka \& Ueta, {\it in prep.}).
As has been discussed above, 
the apparent $c({\rm H}\beta)$ underestimate imposed false reddening in the previous analyses.
Hence, the revised models naturally suggest greater $L_{\ast}$ 
(by a factor of two on average).
Correspondingly, 
the expected initial progenitor mass for the nine-PN sample comes out to be greater.

Especially, the brightest four PNe in M\,31 (M1687, M2068, M2538, and M50) are now appropriately found to be 
the most massive among the low-mass progenitors 
(1.9 -- 2.4\,M$_{\odot}$ with the average of 2.2\,M$_{\odot}$,
as opposed to 1.3 -- 1.6\,M$_{\odot}$ with the average of 1.5\,M$_{\odot}$).
For such relatively higher-mass progenitors, 
we would indeed expect comparatively enhanced N/O abundance ratios.
However, they are by no means N over-abundant as in Type I PN (\citealt{ptp1983}).
Therefore, the present results from fully iterative self-consistent extinction correction and plasma diagnostics
are in reasonable agreement with what is predicted by the existing theoretical models.
There is no need for a greater amount of N from low-mass progenitors.

\section{Concluding Remarks}

Plasma diagnostics determine the physico-chemical conditions of line-emitting objects
in terms of $n_{\rm e}$ and $T_{\rm e}$ as well as ionic and elemental abundances.
Because $c({\rm H}\beta)$ is directly influenced by $n_{\rm e}$ and $T_{\rm e}$, 
$n_{\rm e}$ and $T_{\rm e}$ must be determined self-consistently via 
an iterative search for convergence through both extinction correction and plasma diagnostics.
In the present exercise, 
we have demonstrated, for a sample of nine bright PNe in M\,31, 
how inconsistently assumed $n_{\rm e}$ and $T_{\rm e}$ in extinction correction
can affect the results of the subsequent plasma diagnostics.

If $c({\rm H}\beta)$ is not iteratively updated, 
and given the measured line fluxes,
the resulting $n_{\rm e}$ and $T_{\rm e}$ values are more or less fixed 
as soon as the $n_{\rm e}$- and $T_{\rm e}$-diagnostic lines are selected.
If $c({\rm H}\beta)$ is not determined properly, 
observed line fluxes above and below the reference H$\beta$ wavelength 
can be over/under-corrected depending on their wavelengths.
Consequently, diagnostic line ratios can be erroneously amplified/reduced 
especially when the lines involved are taken from both sides of H$\beta$.
This practically means that the derived ionic and elemental abundances can be unreliable,
even though the resulting $n_{\rm e}$ and $T_{\rm e}$ values may still appear reasonable.
Moreover, the results of model calculations would be equally compromised,
especially when such wrongly extinction-corrected lines are used as constraints.

For the present case of the nine bright PN sample in M\,31, 
we have been able to attribute the differences between the previous and present results 
to $c({\rm H}\beta)$ that was underestimated more than 50\,\% in the previous analyses.
It was because the H$\alpha$-to-H$\beta$ line ratio was assumed to be fixed at 2.86
(equivalent to assuming $n_{\rm e}=10^3$\,cm$^{-3}$ and $T_{\rm e}=10^4$\,K)
in extinction correction even when the final $n_{\rm e}$ and $T_{\rm e}$ were different.
We have also shown that the underestimated $c({\rm H}\beta)$ inflicted inconsistencies 
in the derived ionic and elemental abundances as well as the subsequent photoionization 
model calculations. 
In the end, we have established that this bright M\,31 PN sample represents the high-mass end of 
the low-mass PN progenitor stars of less than solar metallicities.
Hence, no anomalous N overabundance has been found to suspect any irregularities 
in the existing evolutionary models as was hinted at with the previous analyses.

More specifically, the N/O abundance ratio can be significantly affected 
in plasma diagnostics based on optical spectra,
if $c({\rm H}\beta)$ is not determined properly.
For the present case, the previously suspected N overabundance was simply caused 
by the underestimated $c({\rm H}\beta)$.
The empirical N abundance is based almost exclusively on the [N\,{\sc ii}] lines around 6000\,{\AA} 
(5755 and 6548/83\,{\AA}), on the much redder side of the reference wavelength at H$\beta$. 
Therefore, the [N\,{\sc ii}] line strengths, and hence, the N abundance can be artificially inflated 
when $c({\rm H}\beta)$ is underestimated.
In the mean time, the O abundance, based on multiple lines of multiple ionic species 
scattered across the optical spectrum on either side of H$\beta$,
is relatively insensitive to the apparent $c({\rm H}\beta)$ underestimate.
For example, the N/O ratio is often used as an indicator of the initial stellar mass 
based on predictions made by theoretical models (e.g.\ \citealt{kl2016,mb2016,vk2018}).
Hence, the erroneously estimated N/O ratio can lead to a variety of wrong conclusions
beyond the physico-chemical conditions of the target PNe.
This issue is of course not isolated in PNe and can happen in any line-emitting objects.
Therefore, caution must be exercised when quoting abundances from the literature.

To summarize, the general lessons learned are;
\vspace{-6pt}
\begin{itemize}
      \item The quality of the input spectra is of course important. 
      Especially, line fluxes should be free from any anomalies 
      such as the atmospheric dispersion, sky emission/absorption, etc.
      
      \item Extinction correction and plasma diagnostics should be performed simultaneously and self-consistently 
      through full iteration, with a careful choice of diagnostic lines.

      \item It is best to incorporate as many lines as possible in the analyses,
      especially from either side of the reference wavelength of extinction correction (typically at H$\beta$).
      
      \item The adverse effects of not performing extinction correction and plasma diagnostics self-consistently
      do not necessarily incur in the resulting $n_{\rm e}$ and $T_{\rm e}$, but would be more likely 
      in the resulting ionic and elemental abundances via diagnostic line ratios compromised by the wrongly
      derived $c({\rm H}\beta)$.
      
      \item The incorrectly determined $c({\rm H}\beta)$ would systematically redden or blue the input spectrum
      and affect the outcomes of any subsequent analyses. 

      \item Abundances in the literature need to be treated with care, 
      unless plasma diagnostics were performed self-consistently in conjunction with extinction correction.
      It may be worthwhile to re-evaluate abundances via self-consistent extinction correction and plasma diagnostics,
      especially when anomalies are reported in the previous results. 
\end{itemize}

\begin{acknowledgements}
This research made use of 
the NASA/IPAC Infrared Science Archive Galactic Dust Reddening and Extinction (GDRE) service and
PyNeb, a toolset dedicated to the analysis of emission lines \citep{pyneb}.
T.U.\ was supported partially by the Japan Society for the Promotion of Science 
(JSPS) through its invitation fellowship program (FY2020; L20505). 
M.O.\ was supported by JSPS Grants-in-Aid for Scientific Research(C) (JP19K03914 and 22K03675).
Authors thank the anonymous referee, 
whose inputs helped to clarify some critical points in the manuscript.
\end{acknowledgements}

%
%


\begin{appendix} 
\onecolumn 
\section{Supplemental Materials}

\begin{table}[ht]
\centering
\caption{\label{tab:flx1}Extinction-Corrected Line Fluxes and Original-to-Revised Ratios for the Brightest Four PNe in M\,31}
\begin{tabular}{lrcrcrcrcrc}
\hline\hline
\multicolumn{1}{c}{Line} & \multicolumn{1}{c}{M1687}   & Ratio & \multicolumn{1}{c}{M2068}   & Ratio & \multicolumn{1}{c}{M2538}   & Ratio & \multicolumn{1}{c}{M50}   & Ratio \\
\hline
$[$O {\sc ii}$]$\,3727	&	21.0	$\pm$	2.3	&	0.97	&	50.2	$\pm$	9.4	&	0.83	&	66.2	$\pm$	12.1	&	0.77	&	53.4	$\pm$	7.2	&	0.94	\\
H10 3798	&		$\cdots$		&	$\cdots$	&		$\cdots$		&	$\cdots$	&	4.4	$\pm$	0.8	&	0.77	&	5.8	$\pm$	0.8	&	0.94	\\
H9 3835	&		$\cdots$		&	$\cdots$	&	6.1	$\pm$	1.1	&	0.85	&	8.9	$\pm$	1.6	&	0.78	&	5.8	$\pm$	0.8	&	0.94	\\
$[$Ne {\sc iii}$]$\,3868	&	172.7	$\pm$	18.6	&	0.97	&	162.0	$\pm$	30.5	&	0.85	&	119.3	$\pm$	21.8	&	0.79	&	126.3	$\pm$	16.9	&	0.94	\\
H8 3889\tablefootmark{a}	&	0.1	$\pm$	1.6	&	$\cdots$	&	2.6	$\pm$	2.8	&	$\cdots$	&	5.7	$\pm$	2.8	&	$\cdots$	&	2.4	$\pm$	2.0	&	$\cdots$	\\
$[$Ne {\sc iii}$]$\,3967\tablefootmark{b}	&	50.7	$\pm$	6.7	&	0.97	&	49.8	$\pm$	11.3	&	0.81	&	25.3	$\pm$	6.4	&	0.66	&	33.9	$\pm$	5.9	&	0.92	\\
He {\sc i} 4026	&	6.1	$\pm$	0.7	&	0.97	&	2.8	$\pm$	0.5	&	0.87	&	3.8	$\pm$	0.7	&	0.81	&	2.5	$\pm$	0.3	&	0.95	\\
$[$S {\sc ii}$]$\,4069	&	0.0	$\pm$	0.0	&	$\cdots$	&	4.3	$\pm$	0.8	&	0.88	&		$\cdots$		&	$\cdots$	&	3.9	$\pm$	0.5	&	0.96	\\
$[$S {\sc ii}$]$\,4076	&	0.0	$\pm$	0.0	&	$\cdots$	&	2.9	$\pm$	0.5	&	0.88	&		$\cdots$		&	$\cdots$	&	1.1	$\pm$	0.2	&	0.96	\\
H6 4101\tablefootmark{c}	&	29.3	$\pm$	3.1	&	0.98	&	30.2	$\pm$	5.7	&	0.88	&	30.1	$\pm$	5.5	&	0.83	&	27.1	$\pm$	3.6	&	0.95	\\
H5 4340\tablefootmark{d}	&	48.5	$\pm$	5.2	&	0.99	&	50.0	$\pm$	9.4	&	0.92	&	50.7	$\pm$	9.3	&	0.88	&	46.2	$\pm$	6.2	&	0.97	\\
$[$O {\sc iii}$]$\,4363	&	27.0	$\pm$	2.9	&	0.99	&	15.3	$\pm$	2.9	&	0.92	&	19.1	$\pm$	3.5	&	0.88	&	17.7	$\pm$	2.4	&	0.98	\\
He {\sc i} 4472	&	6.0	$\pm$	0.6	&	0.99	&	6.0	$\pm$	1.1	&	0.94	&	4.7	$\pm$	0.9	&	0.91	&	5.1	$\pm$	0.7	&	0.98	\\
He {\sc ii} 4686	&	2.9	$\pm$	0.3	&	1.00	&	3.9	$\pm$	0.7	&	0.97	&	12.5	$\pm$	2.3	&	0.96	&	14.6	$\pm$	1.9	&	0.99	\\
$[$Ar {\sc iv}$]$\,4711\tablefootmark{e}	&	1.9	$\pm$	0.3	&	0.99	&	1.5	$\pm$	0.4	&	0.91	&	2.7	$\pm$	0.6	&	0.93	&	2.4	$\pm$	0.4	&	0.98	\\
$[$Ar {\sc iv}$]$\,4740	&	5.3	$\pm$	0.6	&	1.00	&	4.1	$\pm$	0.8	&	0.98	&	3.4	$\pm$	0.6	&	0.97	&	4.4	$\pm$	0.6	&	0.99	\\
H$\beta$ 4861\tablefootmark{f}	&	100.0	$\pm$	10.5	&	1.00	&	100.0	$\pm$	18.8	&	1.00	&	100.0	$\pm$	18.1	&	1.00	&	100.0	$\pm$	13.1	&	1.00	\\
He {\sc i} 4922	&	0.9	$\pm$	0.1	&	1.01	&	1.4	$\pm$	0.3	&	1.01	&	1.0	$\pm$	0.2	&	1.02	&	0.7	$\pm$	0.1	&	1.01	\\
$[$O {\sc iii}$]$\,4959	&	599.5	$\pm$	76.7	&	1.00	&	506.0	$\pm$	116.2	&	1.02	&	461.1	$\pm$	101.9	&	1.03	&	543.0	$\pm$	86.9	&	1.01	\\
$[$O {\sc iii}$]$\,5007	&	1774.9	$\pm$	227.0	&	1.01	&	1492.0	$\pm$	342.5	&	1.03	&	1360.8	$\pm$	300.6	&	1.04	&	1606.6	$\pm$	256.9	&	1.02	\\
$[$N {\sc i}$]$\,5198	&		$\cdots$		&	$\cdots$	&		$\cdots$		&	$\cdots$	&		$\cdots$		&	$\cdots$	&	0.5	$\pm$	0.1	&	1.01	\\
He {\sc ii} 5411	&		$\cdots$		&	$\cdots$	&		$\cdots$		&	$\cdots$	&	1.4	$\pm$	0.3	&	1.12	&	1.4	$\pm$	0.2	&	1.03	\\
$[$Cl {\sc iii}$]$\,5518	&		$\cdots$		&	$\cdots$	&		$\cdots$		&	$\cdots$	&		$\cdots$		&	$\cdots$	&	0.4	$\pm$	0.1	&	1.04	\\
$[$Cl {\sc iii}$]$\,5538	&		$\cdots$		&	$\cdots$	&		$\cdots$		&	$\cdots$	&		$\cdots$		&	$\cdots$	&	0.8	$\pm$	0.1	&	1.03	\\
$[$N {\sc ii}$]$\,5755	&	2.0	$\pm$	0.3	&	1.02	&	2.3	$\pm$	0.5	&	1.11	&	1.0	$\pm$	0.2	&	1.18	&	2.1	$\pm$	0.3	&	1.04	\\
He {\sc i} 5876	&	16.9	$\pm$	2.1	&	1.02	&	15.7	$\pm$	3.6	&	1.12	&	13.0	$\pm$	2.9	&	1.19	&	15.6	$\pm$	2.5	&	1.04	\\
$[$O {\sc i}$]$\,6300	&	7.5	$\pm$	0.9	&	1.03	&	6.3	$\pm$	1.4	&	1.16	&	5.0	$\pm$	1.1	&	1.25	&	9.1	$\pm$	1.4	&	1.06	\\
$[$S {\sc iii}$]$\,6312\tablefootmark{g}	&	2.1	$\pm$	0.3	&	1.03	&	1.7	$\pm$	0.4	&	1.17	&	1.7	$\pm$	0.4	&	1.26	&	1.6	$\pm$	0.3	&	1.07	\\
$[$O {\sc i}$]$\,6363	&	2.7	$\pm$	0.3	&	1.03	&	2.0	$\pm$	0.5	&	1.16	&	2.0	$\pm$	0.4	&	1.26	&	2.9	$\pm$	0.5	&	1.06	\\
$[$N {\sc ii}$]$\,6548	&	10.7	$\pm$	1.3	&	1.04	&	21.9	$\pm$	5.0	&	1.18	&	12.5	$\pm$	2.7	&	1.29	&	23.3	$\pm$	3.7	&	1.07	\\
H$\alpha$ 6563\tablefootmark{h}	&	277.3	$\pm$	34.6	&	1.03	&	241.7	$\pm$	54.7	&	1.18	&	216.3	$\pm$	47.2	&	1.29	&	268.1	$\pm$	42.1	&	1.07	\\
$[$N {\sc ii}$]$\,6583	&	27.0	$\pm$	3.4	&	1.03	&	63.5	$\pm$	14.4	&	1.18	&	36.3	$\pm$	7.9	&	1.29	&	67.2	$\pm$	10.5	&	1.07	\\
He {\sc i} 6678\tablefootmark{i}	&	3.6	$\pm$	0.5	&	1.03	&	3.7	$\pm$	0.8	&	1.20	&	2.9	$\pm$	0.6	&	1.31	&	3.4	$\pm$	0.5	&	1.07	\\
$[$S {\sc ii}$]$\,6716	&	1.3	$\pm$	0.2	&	1.03	&	1.6	$\pm$	0.4	&	1.20	&	2.0	$\pm$	0.4	&	1.31	&	2.4	$\pm$	0.4	&	1.07	\\
$[$S {\sc ii}$]$\,6730	&	2.3	$\pm$	0.3	&	1.03	&	3.2	$\pm$	0.7	&	1.20	&	3.4	$\pm$	0.7	&	1.31	&	4.5	$\pm$	0.7	&	1.07	\\
$[$Ar V$]$\,7005	&		$\cdots$		&	$\cdots$	&		$\cdots$		&	$\cdots$	&		$\cdots$		&	$\cdots$	&		$\cdots$		&	$\cdots$	\\
He {\sc i} 7065	&	11.0	$\pm$	1.4	&	1.04	&	8.0	$\pm$	1.8	&	1.23	&	6.7	$\pm$	1.5	&	1.36	&	8.7	$\pm$	1.4	&	1.08	\\
$[$Ar {\sc iii}$]$\,7136	&	9.4	$\pm$	1.2	&	1.05	&	13.1	$\pm$	2.9	&	1.23	&	7.6	$\pm$	1.6	&	1.36	&	13.8	$\pm$	2.1	&	1.08	\\
He {\sc i} 7281	&	0.8	$\pm$	0.1	&	1.05	&	0.5	$\pm$	0.1	&	1.25	&	1.1	$\pm$	0.2	&	1.40	&	0.7	$\pm$	0.1	&	1.09	\\
$[$O {\sc ii}$]$\,7320	&	6.6	$\pm$	0.8	&	1.04	&	3.5	$\pm$	0.8	&	1.25	&	2.9	$\pm$	0.6	&	1.40	&	5.8	$\pm$	0.9	&	1.09	\\
$[$O {\sc ii}$]$\,7330	&	6.5	$\pm$	0.8	&	1.04	&	3.4	$\pm$	0.8	&	1.25	&	3.0	$\pm$	0.6	&	1.40	&	5.2	$\pm$	0.8	&	1.09	\\
$[$Ar {\sc iii}$]$\,7751	&	1.8	$\pm$	0.2	&	1.05	&	1.9	$\pm$	0.4	&	1.29	&	1.3	$\pm$	0.3	&	1.47	&	2.5	$\pm$	0.4	&	1.10 \\	
\hline
\end{tabular}
\tablefoot{Contaminating He\,{\sc i} and He\,{\sc ii} lines scaled from He\,{\sc i} 5876 and He\,{\sc ii} 4686, respectively.
\tablefoottext{a}{Blend with He\,{\sc i} 3889.}
\tablefoottext{b}{Blend with H7 3970 and He\,{\sc i} 3965.}
\tablefoottext{c}{Blend with He\,{\sc ii} 4100.}
\tablefoottext{d}{Blend with He\,{\sc ii} 4338.}
\tablefoottext{e}{Blend with He\,{\sc i} 4713.}
\tablefoottext{f}{Blend with He\,{\sc ii} 4859.}
\tablefoottext{g}{Blend with He\,{\sc ii} 6310.}
\tablefoottext{h}{Blend with He\,{\sc ii} 6560.}
\tablefoottext{i}{Blend with He\,{\sc ii} 6683.}}
\end{table}
\FloatBarrier

\begin{table}[ht]
\footnotesize
\centering
\caption{\label{tab:flx2}Extinction-Corrected Line Fluxes and Original-to-Revised Ratios for the Other Five PNe in M\,31}
\begin{tabular}{lrcrcrcrcrc}
\hline\hline
\multicolumn{1}{c}{Line} & \multicolumn{1}{c}{M1596}   & Ratio & \multicolumn{1}{c}{M2471}   & Ratio & \multicolumn{1}{c}{M2860}   & Ratio & \multicolumn{1}{c}{M1074}   & Ratio & \multicolumn{1}{c}{M1675}   & Ratio \\
\hline
$[$O {\sc ii}$]$\,3727	&	65.0$\pm$5.3	&	0.98	&	78.7$\pm$13.7	&	0.80	&	28.7$\pm$7.5	&	0.73	&	20.8$\pm$	1.9	&	0.93	&	17.8$\pm$4.7	&	0.80	\\
H10 3798	            &		$\cdots$	&$\cdots$	&	5.4	$\pm$0.9	&	0.81	&	4.3	$\pm$1.1	&	0.73	&		$\cdots$	&	$\cdots$&	$\cdots$		&	$\cdots$	\\
H9 3835	                &	8.3$\pm$0.7  	&	0.98	&	6.0	$\pm$1.0	&	0.81	&	5.1	$\pm$1.3	&	0.74	&	7.2	$\pm$	0.7	&	0.93	&	5.3	$\pm$1.4	&	0.81	\\
$[$Ne {\sc iii}$]$\,3868&	127.5$\pm$10.4	&	0.98	&	141.9$\pm$24.6	&	0.82	&	115.6$\pm$30.2	&	0.74	&	118.8$\pm$	10.7&	0.93	&	112.4$\pm$29.5	&	0.81	\\
H8 3889\tablefootmark{a}&	0.0$\pm$1.0	    &$\cdots$	&	10.8$\pm$3.1	&$\cdots$	&	5.5	$\pm$4.0	&$\cdots$	&	3.7	$\pm$	1.5	&	$\cdots$&	3.1	$\pm$3.7	&	4.01	\\
$[$Ne {\sc iii}$]$\,3967\tablefootmark{b}
                        &	38.7$\pm$4.0	&	0.97	&	44.4$\pm$9.3	&	0.77	&	45.7$\pm$14.3	&	0.67	&	35.6$\pm$	4.2	&	0.90	&	32.7$\pm$11.0	&	0.73	\\
He {\sc i} 4026	        &	5.6$\pm$0.5	    &	0.98	&	2.7	$\pm$0.5	&	0.84	&	4.6	$\pm$1.2	&	0.78	&	1.6	$\pm$	0.2	&	0.94	&	$\cdots$		&	$\cdots$	\\
$[$S {\sc ii}$]$\,4069	&	3.9$\pm$0.3	    &	0.98	&	3.6	$\pm$0.6	&	0.85	&	3.5	$\pm$0.9	&	0.78	&	2.1	$\pm$	0.2	&	0.94	&	$\cdots$		&	$\cdots$	\\
$[$S {\sc ii}$]$\,4076	&		$\cdots$	&$\cdots$	&	$\cdots$		&$\cdots$	&	1.9	$\pm$0.5	&	0.79	&		$\cdots$	&	$\cdots$&	$\cdots$		&	$\cdots$	\\
H6 4101\tablefootmark{c}&	27.3$\pm$2.2	&	0.99	&	31.3$\pm$5.5	&	0.85	&	32.8$\pm$8.6	&	0.79	&	26.0$\pm$	2.3	&	0.94	&	27.1$\pm$7.2	&	0.85	\\
H5 4340\tablefootmark{d}&	47.0$\pm$3.8	&	0.99	&	50.8$\pm$8.9	&	0.90	&	51.7$\pm$13.7	&	0.85	&	46.2$\pm$	4.1	&	0.96	&	51.7$\pm$13.8	&	0.90	\\
$[$O {\sc iii}$]$\,4363	&	20.1$\pm$1.6	&	0.99	&	18.9$\pm$3.3	&	0.90	&	11.6$\pm$3.1	&	0.86	&	17.5$\pm$	1.6	&	0.96	&	13.9$\pm$3.7	&	0.90	\\
He {\sc i} 4472	        &	4.3$\pm$0.4	    &	0.99	&	4.1$\pm$0.7	    &	0.92	&	6.1	$\pm$1.6	&	0.89	&	5.4	$\pm$	0.5	&	0.97	&	4.7	$\pm$1.3	&	0.92	\\
He {\sc ii} 4686	    &	35.4$\pm$2.8	&	1.00	&	33.6$\pm$5.8	&	0.97	&	7.7	$\pm$2.1	&	0.95	&	2.7	$\pm$	0.2	&	0.99	&	12.8$\pm$3.4	&	0.97	\\
$[$Ar {\sc iv}$]$\,4711\tablefootmark{e}
                        &	6.4$\pm$0.5	    &	1.00	&	4.2$\pm$0.8   	&	0.96	&	3.1	$\pm$0.9	&	0.90	&	1.7	$\pm$	0.2	&	0.96	&	3.3	$\pm$1.0	&	0.94	\\
$[$Ar {\sc iv}$]$\,4740	&	8.1$\pm$0.6	    &	1.00	&	4.8$\pm$0.8	    &	0.98	&	4.5	$\pm$1.2	&	0.97	&	3.8	$\pm$	0.3	&	0.99	&	4.5	$\pm$1.2	&	0.98	\\
H$\beta$ 4861\tablefootmark{f}
                        &	100.0$\pm$7.7	&	1.00	&	100.0$\pm$17.0	&	1.00	&	100.0$\pm$26.7	&	1.00	&	100.0$\pm$	8.7	&	1.00	&	100.0$\pm$26.7	&	1.00	\\
He {\sc i} 4922	        &		$\cdots$	&$\cdots$	&	0.7$\pm$0.2	    &	1.03	&	1.3	$\pm$0.4	&	1.02	&		$\cdots$	&	$\cdots$&	2.4	$\pm$0.8	&	1.02	\\
$[$O {\sc iii}$]$\,4959	&	575.3$\pm$54.3	&	1.02	&	499.8$\pm$103.7	&	1.03	&	477.6$\pm$156.1	&	1.03	&	476.0$\pm$	50.8&	1.01	&	510.6$\pm$167.0	&	1.02	\\
$[$O {\sc iii}$]$\,5007	&	1717.7$\pm$161.9&	1.01	&	1491.1$\pm$309.3&	1.04	&	1411.8$\pm$461.7&	1.04	&   1429.6$\pm$152.4&	1.01	& 1535.9$\pm$502.5	&	1.03	\\
$[$N {\sc i}$]$\,5198	&		$\cdots$	&$\cdots$	&		$\cdots$	&$\cdots$	&		$\cdots$	&$\cdots$	&	$\cdots$		&	$\cdots$&	2.1	$\pm$0.7	&	1.07	\\
He {\sc ii} 5411	    &	3.1$\pm$0.3	    &	1.03	&	2.6$\pm$0.5	    &	1.11	&	0.5	$\pm$0.2	&	1.15	&	$\cdots$		&	$\cdots$&		$\cdots$	&	$\cdots$	\\
$[$Cl {\sc iii}$]$\,5518&	0.6$\pm$0.1	    &	1.02	&		$\cdots$	&$\cdots$	&	0.3	$\pm$0.1	&	1.16	&	$\cdots$		&	$\cdots$&		$\cdots$	&	$\cdots$	\\
$[$Cl {\sc iii}$]$\,5538&	1.2$\pm$0.1  	&	1.03	&		$\cdots$	&$\cdots$	&	0.4	$\pm$0.1	&	1.18	&	$\cdots$		&	$\cdots$&		$\cdots$	&	$\cdots$	\\
$[$N {\sc ii}$]$\,5755	&	2.5$\pm$0.2	    &	1.03	&	0.9	$\pm$0.2	&	1.16	&	0.8	$\pm$0.3	&	1.21	&	0.5	$\pm$	0.1	&	1.05	&	2.3	$\pm$0.8	&	1.15	\\
He {\sc i} 5876     	&	14.1$\pm$1.3 	&	1.03	&	10.9$\pm$2.2	&	1.17	&	13.5$\pm$4.4	&	1.23	&	16.1$\pm$	1.7	&	1.05	&	13.9$\pm$4.6	&	1.16	\\
$[$O {\sc i}$]$\,6300	&	7.0$\pm$0.6	    &	1.03	&	6.6	$\pm$1.4	&	1.22	&	2.2	$\pm$0.7	&	1.31	&	3.6	$\pm$	0.4	&	1.07	&	7.7	$\pm$2.5	&	1.20	\\
$[$S {\sc iii}$]$\,6312\tablefootmark{g}
                        &	4.0$\pm$0.4	    &	1.03	&	1.3	$\pm$0.3	&	1.24	&	1.3	$\pm$0.4	&	1.32	&	2.1	$\pm$	0.2	&	1.07	&	2.2	$\pm$0.7	&	1.22	\\
$[$O {\sc i}$]$\,6363	&	2.2$\pm$0.2	    &	1.04	&	2.0	$\pm$0.4	&	1.22	&	0.9	$\pm$0.3	&	1.32	&	1.2	$\pm$	0.1	&	1.07	&	2.4	$\pm$0.8	&	1.22	\\
$[$N {\sc ii}$]$\,6548	&	36.4$\pm$3.3	&	1.04	&	14.7$\pm$3.0	&	1.24	&	9.0	$\pm$3.0	&	1.35	&	5.9	$\pm$	0.6	&	1.08	&	29.0$\pm$9.5	&	1.24	\\
H$\alpha$ 6563\tablefootmark{h}
                        &	276.2$\pm$25.5	&	1.04	&	229.0$\pm$47.0	&	1.25	&	211.0$\pm$69.2	&	1.36	&	265.3$\pm$	27.6	&	1.08&	230.3$\pm$75.6	&	1.24	\\
$[$N {\sc ii}$]$\,6583	&	105.8$\pm$9.7	&	1.04	&	42.4$\pm$8.6	&	1.25	&	24.8$\pm$8.1	&	1.36	&	17.1$\pm$	1.8	&	1.08	&	84.9$\pm$27.8	&	1.24	\\
He {\sc i} 6678\tablefootmark{i}
                        &	3.8$\pm$0.4	    &	1.04	&	2.2	$\pm$0.5	&	1.29	&	2.9	$\pm$0.9	&	1.39	&	3.5	$\pm$	0.4	&	1.08	&	4.2	$\pm$1.4	&	1.26	\\
$[$S {\sc ii}$]$\,6716	&	6.2$\pm$0.6	    &	1.04	&	2.6	$\pm$0.5	&	1.27	&	1.5	$\pm$0.5	&	1.38	&	1.2	$\pm$	0.1	&	1.08	&	3.4	$\pm$1.1	&	1.26	\\
$[$S {\sc ii}$]$\,6730	&	10.6$\pm$1.0	&	1.03	&	4.4	$\pm$0.9	&	1.27	&	2.6	$\pm$0.8	&	1.39	&	1.9	$\pm$	0.2	&	1.08	&	7.1	$\pm$2.3	&	1.26	\\
$[$Ar V$]$\,7005	    &	1.0$\pm$0.1	    &	1.04	&	0.4	$\pm$0.1	&	1.30	&	$\cdots$		&$\cdots$	&	$\cdots$		&	$\cdots$&		$\cdots$	&	$\cdots$	\\
He {\sc i} 7065       	&	6.7$\pm$0.6	    &	1.04	&	5.2	$\pm$1.1	&	1.30	&	5.1	$\pm$1.7	&	1.45	&	8.5	$\pm$	0.9	&	1.09	&	5.5	$\pm$1.8	&	1.30	\\
$[$Ar {\sc iii}$]$\,7136&	22.4$\pm$2.0	&	1.04	&	9.5	$\pm$1.9	&	1.31	&	9.9	$\pm$3.2	&	1.46	&	7.6	$\pm$	0.8	&	1.10	&	11.3$\pm$3.7	&	1.31	\\
He {\sc i} 7281	        &		$\cdots$	&$\cdots$	&	0.5	$\pm$0.1	&	1.32	&	0.3	$\pm$0.1	&	1.51	&	$\cdots$		&	$\cdots$&		$\cdots$	&	$\cdots$	\\
$[$O {\sc ii}$]$\,7320	&	4.4$\pm$0.4	    &	1.04	&	3.4	$\pm$0.7	&	1.34	&	1.5	$\pm$0.5	&	1.50	&	3.7	$\pm$	0.4	&	1.10	&		$\cdots$	&	$\cdots$	\\
$[$O {\sc ii}$]$\,7330	&	4.0$\pm$0.4	    &	1.04	&	3.1	$\pm$0.6	&	1.34	&	1.4	$\pm$0.5	&	1.50	&	3.7	$\pm$	0.4	&	1.10	&	2.6	$\pm$0.8	&	1.33	\\
$[$Ar {\sc iii}$]$\,7751&	4.2$\pm$0.4	    &	1.05	&	2.0	$\pm$0.4	&	1.39	&	1.8	$\pm$0.6	&	1.59	&	1.4	$\pm$	0.2	&	1.12	&	2.2	$\pm$0.7	&	1.43	\\
\hline
\end{tabular}
\tablefoot{Contaminating He\,{\sc i} and He\,{\sc ii} lines scaled from He\,{\sc i} 5876 and He\,{\sc ii} 4686, respectively.
\tablefoottext{a}{Blend with He\,{\sc i} 3889.}
\tablefoottext{b}{Blend with H7 3970 and He\,{\sc i} 3965.}
\tablefoottext{c}{Blend with He\,{\sc ii} 4100.}
\tablefoottext{d}{Blend with He\,{\sc ii} 4338.}
\tablefoottext{e}{Blend with He\,{\sc i} 4713.}
\tablefoottext{f}{Blend with He\,{\sc ii} 4859.}
\tablefoottext{g}{Blend with He\,{\sc ii} 6310.}
\tablefoottext{h}{Blend with He\,{\sc ii} 6560.}
\tablefoottext{i}{Blend with He\,{\sc ii} 6683.}}
\end{table}

\begin{table*}
\centering
\caption{\label{tab:nT}Derived $n_{\rm e}$ and $T_{\rm e}$ for both high- and low-excitation regions 
from fully iterative extinction correction/plasma diagnostics with 
original-to-revised ratios.}
\begin{tabular}{lcccccc} 
\hline\hline
& M1687 & Ratio & M2068 & Ratio & M2538 & Ratio\\
\hline
$T_{\rm e}$([{\rm O}\,{\sc iii}]) & 
12940$\pm$\phantom{1}420 &0.983& 
11250$\pm$\phantom{1}310 &0.945& 
12870$\pm$\phantom{1}550 &0.942\\
$n_{\rm e}$([{\rm Ar}\,{\sc iv}]) & 
22340$\pm$3510 &1.386& 
18020$\pm$2530 &1.970&  
\phantom{1}4430$\pm$1230 &1.563\\
$T_{\rm e}$([{\rm N}\,{\sc ii}])  & 
25650$\pm$1690 &0.842& 
13020$\pm$1110 &0.795& 
12600$\pm$\phantom{1}580  &0.947\\
$n_{\rm e}$([{\rm S}\,{\sc ii}])  & 
\phantom{1}4720$\pm$1320 &0.977& 
11010$\pm$5120 & 1.771&  
\phantom{1}3340$\pm$\phantom{1}880  &1.002\\
\hline\hline
& M50 & Ratio & M1596 & Ratio & M2471 & Ratio \\
\hline
$T_{\rm e}$([{\rm O}\,{\sc iii}]) & 
11670$\pm$\phantom{1}320 &0.984& 
12030$\pm$\phantom{1}340 &0.992& 
12380$\pm$\phantom{1}380 &0.954\\
$n_{\rm e}$([{\rm Ar}\,{\sc iv}]) & 
\phantom{1}9780$\pm$1690 &1.364&  
\phantom{1}5410$\pm$1100 &1.116&  
\phantom{1}3640$\pm$\phantom{1}930 &1.312\\
$T_{\rm e}$([{\rm N}\,{\sc ii}])  & 
13030$\pm$\phantom{1}670  &0.979& 
11770$\pm$\phantom{1}450  &0.994& 
11390$\pm$\phantom{1}420  &0.959\\
$n_{\rm e}$([{\rm S}\,{\sc ii}])  & 
\phantom{1}5980$\pm$2120  &1.081&  
\phantom{1}3460$\pm$1030  &1.019&  
\phantom{1}3220$\pm$\phantom{1}870  &1.000\\
\hline\hline
& M2860 & Ratio & M1074 & Ratio & M1675 & Ratio \\
\hline
$T_{\rm e}$([{\rm O}\,{\sc iii}]) & 
10660$\pm$\phantom{1}310 &0.939& 
12110$\pm$\phantom{1}340 &0.976& 
11000$\pm$\phantom{1}270  &0.959\\
$n_{\rm e}$([{\rm Ar}\,{\sc iv}]) & 
\phantom{1}6000$\pm$1070 &1.527& 
15590$\pm$2480 &1.438& 
\phantom{1}5170$\pm$1050 &1.720\\
$T_{\rm e}$([{\rm N}\,{\sc ii}])  & 
13780$\pm$\phantom{1}560  &0.910& 
13120$\pm$\phantom{1}480  &0.977& 
11380$\pm$\phantom{1}860  &1.002\\
$n_{\rm e}$([{\rm S}\,{\sc ii}])  &  
\phantom{1}3440$\pm$\phantom{1}890  &0.994&  
\phantom{1}2750$\pm$\phantom{1}710  &1.006& 
10330$\pm$4680 &2.351\\
\hline
\end{tabular}
\end{table*}

\longtab[4]{
\begin{landscape}
\begin{longtable}{lrcrcrcrc}
\caption{\label{tab:ion1}Ionic Abundances and Original-to-Revised Ratios for the Brightest Four PNe in M\,31}\\
\hline\hline
\multicolumn{1}{c}{Line/Ion}&\multicolumn{1}{c}{M1678} & Ratio & \multicolumn{1}{c}{M2068}      & Ratio & \multicolumn{1}{c}{M2538}      & Ratio & \multicolumn{1}{c}{M50}      & Ratio \\
\hline
\endfirsthead
\caption{continued.}\\
\hline\hline
\multicolumn{1}{c}{Line/Ion}&\multicolumn{1}{c}{M1678} & Ratio & \multicolumn{1}{c}{M2068}      & Ratio & \multicolumn{1}{c}{M2538}      & Ratio & \multicolumn{1}{c}{M50}      & Ratio \\
\hline
\endhead
\hline
\endfoot
He\,{\sc I} 4026        &20.763$\pm$2.247($-2$) &  & 11.255$\pm$2.102($-2$) &  & 15.521$\pm$2.767($-2$) &  & 10.098$\pm$1.378($-2$) &  \\
He\,{\sc I} 4472        & 8.840$\pm$0.978($-2$) &  & 11.040$\pm$2.065($-2$) &  & 8.908$\pm$1.617($-2$) &  & 9.438$\pm$1.258($-2$) &  \\
He\,{\sc } 4922         & 4.428$\pm$0.650($-2$) &  & 9.240$\pm$2.113($-2$) &  & 7.155$\pm$1.589($-2$) &  & 4.986$\pm$0.860($-2$) &  \\
He\,{\sc I} 5876        & 6.472$\pm$0.862($-2$) &  & 8.701$\pm$2.094($-2$) &  & 8.093$\pm$1.820($-2$) &  & 9.086$\pm$1.473($-2$) &  \\
He\,{\sc I} 6678        & 4.779$\pm$0.660($-2$) &  & 7.319$\pm$1.766($-2$) &  & 6.457$\pm$1.423($-2$) &  & 7.084$\pm$1.193($-2$) &  \\
He\,{\sc I} 7065        &11.290$\pm$1.567($-2$) &  & 12.201$\pm$3.179($-2$) &  & 13.152$\pm$3.011($-2$) &  & 14.545$\pm$2.673($-2$) &  \\
He\,{\sc I} 7281        & 4.671$\pm$0.663($-2$) &  & 3.841$\pm$0.982($-2$) &  & 9.850$\pm$2.139($-2$) &  & 6.262$\pm$1.130($-2$) &  \\
\hline
He$^{+}$(adopted)       & 6.472$\pm$0.862($-2$) &1.45 & 8.701$\pm$2.094($-2$) &1.22 & 8.093$\pm$1.820($-2$) &1.14 & 9.086$\pm$1.473($-2$) & 1.08 \\
\hline
\hline
He\,{\sc II} 4686       & 2.513$\pm$0.271($-3$) &  & 3.299$\pm$0.603($-3$) &  & 1.071$\pm$0.191($-2$) &  & 1.235$\pm$0.160($-2$) &  \\
He\,{\sc II} 5411       & $\cdots$        &  & $\cdots$        &  & 1.525$\pm$0.331($-2$) &  & 1.478$\pm$0.235($-2$) &  \\
\hline
He$^{2+}$(adopted)      & 2.513$\pm$0.271($-3$) &0.99 & 3.299$\pm$0.603($-3$) &0.97 & 1.071$\pm$0.191($-2$) &0.95 & 1.235$\pm$0.160($-2$) & 0.98 \\
\hline
\hline
$[$O\,{\sc I}$]$ 6300   & 1.001$\pm$0.177($-6$) &  & 4.826$\pm$1.636($-6$) &  & 4.247$\pm$1.121($-6$) &  & 7.030$\pm$1.587($-6$) &  \\
$[$O\,{\sc I}$]$ 6363   & 1.144$\pm$0.197($-6$) &  & 4.964$\pm$1.675($-6$) &  & 5.216$\pm$1.361($-6$) &  & 7.021$\pm$1.536($-6$) &  \\
\hline
O$^{0}$(adopted)        & 1.064$\pm$0.265($-6$) &  & 4.893$\pm$2.342($-6$) &  & 4.639$\pm$1.763($-6$) &  & 7.025$\pm$2.208($-6$) &  \\
\hline
\hline
$[$O\,{\sc II}$]$ 3727 & 1.180$\pm$0.228($-5$) &  & 4.077$\pm$0.961($-5$) &  & 1.516$\pm$0.377($-5$) &  & 2.541$\pm$0.494($-5$) &  \\
$[$O\,{\sc II}$]$ 7320 & 1.497$\pm$0.288($-5$) &  & 1.644$\pm$0.435($-5$) &  & 1.025$\pm$0.317($-5$) &  & 2.614$\pm$0.564($-5$) &  \\
$[$O\,{\sc II}$]$ 7330 & 1.819$\pm$0.359($-5$) &  & 1.944$\pm$0.539($-5$) &  & 1.309$\pm$0.403($-5$) &  & 2.882$\pm$0.625($-5$) &  \\
\hline
O$^{+}$(adopted)        & 1.405$\pm$0.514($-5$) &1.50 & 2.017$\pm$1.185($-5$) &2.43 & 1.250$\pm$0.637($-5$) &1.82 & 2.653$\pm$0.976($-5$) & 0.74 \\
\hline
\hline
$[$O\,{\sc III}$]$4363 & 2.830$\pm$0.552($-4$) &  & 3.582$\pm$0.881($-4$) &  & 2.168$\pm$0.618($-4$) &  & 3.457$\pm$0.721($-4$) &  \\
$[$O\,{\sc III}$]$4959 & 2.841$\pm$0.440($-4$) &  & 3.621$\pm$0.904($-4$) &  & 2.164$\pm$0.544($-4$) &  & 3.411$\pm$0.612($-4$) &  \\
$[$O\,{\sc III}$]$5007 & 2.818$\pm$0.439($-4$) &  & 3.554$\pm$0.852($-4$) &  & 2.123$\pm$0.533($-4$) &  & 3.377$\pm$0.600($-4$) &  \\
\hline
O$^{2+}$(adopted)       & 2.830$\pm$0.832($-4$) &1.08 & 3.584$\pm$1.523($-4$) &1.26 & 2.150$\pm$0.981($-4$) &1.25 & 3.410$\pm$1.120($-4$) & 1.09 \\
\hline
\hline
$[$Ne\,{\sc III}$]$3868& 7.234$\pm$1.091($-5$) &  & 1.100$\pm$0.235($-4$) &  & 5.093$\pm$1.191($-5$) &  & 7.525$\pm$1.266($-5$) &  \\
$[$Ne\,{\sc III}$]$3967& 7.026$\pm$1.165($-5$) &  & 1.133$\pm$0.279($-4$) &  & 3.599$\pm$1.022($-5$) &  & 6.687$\pm$1.298($-5$) &  \\
\hline
Ne$^{2+}$(adopted)      & 7.137$\pm$1.597($-5$) &1.05 & 1.114$\pm$0.365($-4$) &1.05 & 4.232$\pm$1.570($-5$) &0.15 & 7.116$\pm$1.813($-5$) & 1.08 \\
\hline
\hline
$[$Ar\,{\sc III}$]$7136& 4.436$\pm$0.602($-7$) &  & 8.313$\pm$1.879($-7$) &  & 3.605$\pm$0.828($-7$) &  & 8.087$\pm$1.278($-7$) &  \\
$[$Ar\,{\sc III}$]$7751& 3.627$\pm$0.505($-7$) &  & 4.842$\pm$1.105($-7$) &  & 2.637$\pm$0.603($-7$) &  & 6.036$\pm$0.967($-7$) &  \\
\hline
Ar$^{2+}$(adopted)      & 3.962$\pm$0.786($-7$) &1.21 & 8.313$\pm$1.879($-7$) &1.38 & 2.973$\pm$1.024($-7$) &1.88 & 6.783$\pm$1.603($-7$) & 1.35 \\
\hline
\hline
$[$Ar\,{\sc IV}$]$4711 & 4.329$\pm$0.791($-7$) &  & 4.741$\pm$1.307($-7$) &  & 3.771$\pm$0.904($-7$) &  & 5.584$\pm$1.046($-7$) &  \\
$[$Ar\,{\sc IV}$]$4740 & 5.503$\pm$0.757($-7$) &  & 6.534$\pm$1.298($-7$) &  & 4.504$\pm$1.001($-7$) &  & 6.824$\pm$1.100($-7$) &  \\
\hline
Ar$^{3+}$(adopted)      & 4.942$\pm$1.095($-7$) &1.13 & 5.644$\pm$1.842($-7$) &1.29 & 4.100$\pm$1.349($-7$) &1.18 & 6.173$\pm$1.518($-7$) & 1.10 \\
\hline
\hline
$[$N\,{\sc I}$]$ 5198  & $\cdots$         &  & $\cdots$        &  & $\cdots$         &  & 5.038$\pm$1.593($-7$) &  \\
\hline
N$^{0}$(adopted)        & $\cdots$        &  & $\cdots$        &  & $\cdots$        &  & 5.038$\pm$1.593($-7$) &  \\
\hline
\hline
$[$N\,{\sc II}$]$5755  & 0.921$\pm$0.177($-6$) &  & 7.854$\pm$3.040($-6$) &  & 4.302$\pm$1.291($-6$) &  & 7.741$\pm$1.942($-6$) &  \\
$[$N\,{\sc II}$]$6548  & 1.089$\pm$0.169($-6$) &  & 7.851$\pm$2.317($-6$) &  & 4.440$\pm$1.024($-6$) &  & 7.905$\pm$1.503($-6$) &  \\
$[$N\,{\sc II}$]$6583  & 9.275$\pm$1.463($-7$) &  & 7.631$\pm$2.234($-6$) &  & 4.381$\pm$1.038($-6$) &  & 7.743$\pm$1.447($-6$) &  \\
\hline
N$^{+}$(adopted)        & 9.752$\pm$2.853($-7$) &3.42 & 7.763$\pm$4.427($-6$) &2.55 & 4.385$\pm$1.948($-6$) &1.48 & 7.803$\pm$2.850($-6$) & 1.12 \\
\hline
\hline
$[$S\,{\sc II}$]$4069  & $\cdots$         &  & 2.395$\pm$0.827($-7$) &  & $\cdots$        &  & 1.631$\pm$0.346($-7$) &  \\
$[$S\,{\sc II}$]$4076  & $\cdots$        &  & 9.046$\pm$2.852($-7$) &  & $\cdots$        &  & 3.738$\pm$0.835($-7$) &  \\
$[$S\,{\sc II}$]$6716  & 2.689$\pm$0.596($-8$) &  & 1.608$\pm$0.690($-7$) &  & 1.137$\pm$0.309($-7$) &  & 1.684$\pm$0.472($-7$) &  \\
$[$S\,{\sc II}$]$6730  & 3.016$\pm$0.507($-8$) &  & 1.778$\pm$0.669($-7$) &  & 1.263$\pm$0.306($-7$) &  & 1.873$\pm$0.425($-7$) &  \\
\hline
S$^{+}$(adopted)        & 2.879$\pm$0.783($-8$) &3.72 & 1.695$\pm$0.961($-7$) &4.86 & 1.201$\pm$0.435($-7$) &1.70 & 1.788$\pm$0.635($-7$) & 1.40 \\
\hline
\hline
$[$S\,{\sc III}$]$6312 & 1.607$\pm$0.271($-6$) &  & 2.212$\pm$0.547($-6$) &  & 1.410$\pm$0.372($-6$) &  & 1.791$\pm$0.338($-6$) &  \\
\hline
S$^{2+}$(adopted)       & 1.607$\pm$0.271($-6$) &1.08 & 2.212$\pm$0.547($-6$) &1.42 & 1.410$\pm$0.372($-6$) &1.55 & 1.791$\pm$0.338($-6$) & 1.14 \\
\hline
\hline
$[$Cl\,{\sc III}$]$5518& $\cdots$        &  & $\cdots$        &  & $\cdots$        &  & 2.890$\pm$0.612($-8$) &  \\
$[$Cl\,{\sc III}$]$5538& $\cdots$         &  & $\cdots$        &  & $\cdots$        &  & 5.903$\pm$1.065($-8$) &  \\
\hline
Cl$^{2+}$(adopted)      & $\cdots$        &  & $\cdots$        &  & $\cdots$        &  & 5.903$\pm$1.065($-8$) & 1.10 \\
\hline
\hline
\end{longtable}
\tablefoot{The adopted notation of the abundance (X$^{i+}$/H$^{+}$), $\alpha\pm\beta(-\gamma)$, means $(a \pm b) \times 10^{-\gamma}$.} 
\end{landscape}
}

\longtab[5]{
\begin{landscape}
\begin{longtable}{lrcrcrcrcrc}
\caption{\label{tab:ion2}Ionic Abundances and Original-to-Revised Ratios for the Other Five PNe in M\,31}\\
\hline\hline
\multicolumn{1}{c}{Line/Ion} & \multicolumn{1}{c}{M1596}      & Ratio & \multicolumn{1}{c}{M2471}      & Ratio & \multicolumn{1}{c}{M2860}      & Ratio & \multicolumn{1}{c}{M1074}      & Ratio & \multicolumn{1}{c}{M1675}      & Ratio \\
\hline
\endfirsthead
\caption{continued.}\\
\hline\hline
\multicolumn{1}{c}{Line/Ion} & \multicolumn{1}{c}{M1596}      & Ratio & \multicolumn{1}{c}{M2471}      & Ratio & \multicolumn{1}{c}{M2860}      & Ratio & \multicolumn{1}{c}{M1074}      & Ratio & \multicolumn{1}{c}{M1675}      & Ratio \\
\hline
\endhead
\hline
\endfoot
He\,{\sc I} 4026         & 23.120$\pm$1.809($-2$)  &    & 11.248$\pm$1.879($-2$)  &    & 18.550$\pm$4.793($-2$)  &    &  6.538$\pm$0.640($-2$)  &    &  $\cdots$               &    \\
He\,{\sc I} 4472         &  8.334$\pm$0.666($-2$)  &    &  8.003$\pm$1.358($-2$)  &    & 11.377$\pm$3.056($-2$)  &    & 10.259$\pm$0.917($-2$)  &    &  8.809$\pm$2.349($-2$)  &    \\
He\,{\sc } 4922          &  $\cdots$               &    &  5.163$\pm$1.124($-2$)  &    &  9.027$\pm$2.900($-2$)  &    &  $\cdots$               &    & 16.353$\pm$5.232($-2$)  &    \\
He\,{\sc I} 5876         &  8.992$\pm$0.858($-2$)  &    &  7.056$\pm$1.429($-2$)  &    &  7.953$\pm$2.537($-2$)  &    &  9.992$\pm$1.134($-2$)  &    &  8.478$\pm$2.789($-2$)  &    \\
He\,{\sc I} 6678         &  8.691$\pm$0.829($-2$)  &    &  4.976$\pm$1.059($-2$)  &    &  6.145$\pm$1.971($-2$)  &    &  7.610$\pm$0.811($-2$)  &    &  9.243$\pm$3.021($-2$)  &    \\
He\,{\sc I} 7065         & 14.203$\pm$1.721($-2$)  &    & 11.479$\pm$2.509($-2$)  &    &  8.858$\pm$2.903($-2$)  &    & 16.511$\pm$2.064($-2$)  &    &  9.937$\pm$3.446($-2$)  &    \\
He\,{\sc I} 7281         &  $\cdots$               &    &  5.235$\pm$1.116($-2$)  &    &  2.179$\pm$0.755($-2$)  &    &  $\cdots$               &    &  $\cdots$               &    \\
\hline
He$^{+}$(adopted)        &  8.992$\pm$0.858($-2$)  &0.99&  7.056$\pm$1.429($-2$)  &1.13&  7.953$\pm$2.537($-2$)  &1.33&  9.992$\pm$1.134($-2$)  &0.98&  8.478$\pm$2.789($-2$)  &  1.19  \\
\hline
\hline
He\,{\sc II} 4686        &  2.993$\pm$0.232($-2$)  &    &  2.887$\pm$0.495($-2$)  &    &  6.490$\pm$1.687($-3$)  &    &  2.260$\pm$0.199($-3$)  &    &  1.063$\pm$0.283($-2$)  &    \\
He\,{\sc II} 5411        &  3.323$\pm$0.309($-2$)  &    &  2.843$\pm$0.581($-2$)  &    &  5.745$\pm$1.910($-3$)  &    &  $\cdots$               &    &  $\cdots$               &    \\
\hline
He$^{2+}$(adopted)       &  2.993$\pm$0.232($-2$)  &1.00&  2.887$\pm$0.495($-2$)  &0.95&  6.490$\pm$1.687($-3$)  &0.94&  2.260$\pm$0.199($-3$)  &0.99&  1.063$\pm$0.283($-2$)  &  0.97  \\
\hline
\hline
$[$O\,{\sc I}$]$6300     &  7.494$\pm$1.150($-6$)  &    &  7.906$\pm$1.885($-6$)  &    &  1.429$\pm$0.513($-6$)  &    &  2.691$\pm$0.410($-6$)  &    &  9.044$\pm$3.703($-6$)  &    \\
$[$O\,{\sc I}$]$6363     &  7.230$\pm$1.135($-6$)  &    &  7.610$\pm$1.823($-6$)  &    &  1.873$\pm$0.655($-6$)  &    &  2.801$\pm$0.423($-6$)  &    &  8.722$\pm$3.592($-6$)  &    \\
\hline
O$^{0}$(adopted)         &  7.360$\pm$1.615($-6$)  &    &  7.753$\pm$2.622($-6$)  &    &  1.598$\pm$0.832($-6$)  &    &  2.745$\pm$0.589($-6$)  &    &  8.878$\pm$5.159($-6$)  &    \\
\hline
\hline
$[$O\,{\sc II}$]$3727    &  2.064$\pm$0.325($-5$)  &    &  1.922$\pm$0.415($-5$)  &    &  1.512$\pm$0.446($-5$)  &    &  1.154$\pm$0.200($-5$)  &    &  0.778$\pm$0.225($-5$)  &    \\
$[$O\,{\sc II}$]$7320    &  2.025$\pm$0.361($-5$)  &    &  1.529$\pm$0.397($-5$)  &    &  1.254$\pm$0.473($-5$)  &    &  1.217$\pm$0.208($-5$)  &    &  $\cdots$               &    \\
$[$O\,{\sc II}$]$7330    &  2.320$\pm$0.413($-5$)  &    &  1.777$\pm$0.466($-5$)  &    &  1.492$\pm$0.551($-5$)  &    &  1.547$\pm$0.268($-5$)  &    &  2.366$\pm$0.868($-5$)  &    \\
\hline
O$^{+}$(adopted)         &  2.116$\pm$0.637($-5$)  &1.28&  1.733$\pm$0.740($-5$)  &2.23&  1.417$\pm$0.852($-5$)  &0.63&  1.266$\pm$0.393($-5$)  &1.74&  2.366$\pm$0.868($-5$)  &  0.63  \\
\hline
\hline
$[$O\,{\sc III}$]$4363   &  3.326$\pm$0.576($-4$)  &    &  2.672$\pm$0.632($-4$)  &    &  3.954$\pm$1.247($-4$)  &    &  2.716$\pm$0.469($-4$)  &    &  3.928$\pm$1.205($-4$)  &    \\
$[$O\,{\sc III}$]$4959   &  3.298$\pm$0.410($-4$)  &    &  2.629$\pm$0.597($-4$)  &    &  4.004$\pm$1.357($-4$)  &    &  2.708$\pm$0.356($-4$)  &    &  3.805$\pm$1.263($-4$)  &    \\
$[$O\,{\sc III}$]$5007   &  3.296$\pm$0.405($-4$)  &    &  2.618$\pm$0.588($-4$)  &    &  3.948$\pm$1.342($-4$)  &    &  2.722$\pm$0.357($-4$)  &    &  3.878$\pm$1.260($-4$)  &    \\
\hline
O$^{2+}$(adopted)        &  3.303$\pm$0.815($-4$)  &1.05&  2.638$\pm$1.049($-4$)  &1.20&  3.968$\pm$2.280($-4$)  &1.29&  2.715$\pm$0.688($-4$)  &1.10&  3.872$\pm$2.153($-4$)  &  1.17  \\
\hline
\hline
$[$Ne\,{\sc III}$]$3868  &  6.843$\pm$0.868($-5$)  &    &  6.891$\pm$1.385($-5$)  &    &  0.967$\pm$0.273($-4$)  &    &  6.223$\pm$0.795($-5$)  &    &  8.376$\pm$2.335($-5$)  &    \\
$[$Ne\,{\sc III}$]$3967  &  6.938$\pm$0.977($-5$)  &    &  7.092$\pm$1.630($-5$)  &    &  1.269$\pm$0.421($-4$)  &    &  6.190$\pm$0.923($-5$)  &    &  7.969$\pm$2.705($-5$)  &    \\
\hline
Ne$^{2+}$(adopted)       &  6.885$\pm$1.308($-5$)  &1.02&  6.975$\pm$2.139($-5$)  &0.96&  1.057$\pm$0.502($-4$)  &0.88&  6.209$\pm$1.218($-5$)  &1.02&  8.202$\pm$3.574($-5$)  &  0.98  \\
\hline
\hline
$[$Ar\,{\sc III}$]$7136  &  1.231$\pm$0.132($-6$)  &    &  4.932$\pm$1.010($-7$)  &    &  7.022$\pm$2.328($-7$)  &    &  4.079$\pm$0.475($-7$)  &    &  7.454$\pm$2.496($-7$)  &    \\
$[$Ar\,{\sc III}$]$7751  &  0.954$\pm$0.102($-6$)  &    &  4.241$\pm$0.890($-7$)  &    &  5.269$\pm$1.717($-7$)  &    &  3.221$\pm$0.381($-7$)  &    &  6.064$\pm$1.920($-7$)  &    \\
\hline
Ar$^{2+}$(adopted)       &  1.057$\pm$0.167($-6$)  &1.24&  4.543$\pm$1.346($-7$)  &1.58&  5.887$\pm$2.892($-7$)  &2.04&  3.557$\pm$0.610($-7$)  &1.34&  6.581$\pm$3.149($-7$)  &  1.64  \\
\hline
\hline
$[$Ar\,{\sc IV}$]$4711   &  1.139$\pm$0.141($-6$)  &    &  6.364$\pm$1.318($-7$)  &    &  0.819$\pm$0.259($-6$)  &    &  3.970$\pm$0.627($-7$)  &    &  7.553$\pm$2.353($-7$)  &    \\
$[$Ar\,{\sc IV}$]$4740   &  1.253$\pm$0.149($-6$)  &    &  7.174$\pm$1.356($-7$)  &    &  1.003$\pm$0.278($-6$)  &    &  5.007$\pm$0.610($-7$)  &    &  9.099$\pm$2.519($-7$)  &    \\
\hline
Ar$^{3+}$(adopted)       &  1.193$\pm$0.205($-6$)  &1.04&  6.758$\pm$1.891($-7$)  &1.14&  0.905$\pm$0.380($-6$)  &1.21&  4.503$\pm$0.875($-7$)  &1.13&  8.273$\pm$3.447($-7$)  &  1.15  \\
\hline
\hline
$[$Ar\,{\sc V}$]$7005    &  1.258$\pm$0.152($-7$)  &    &  4.875$\pm$1.115($-8$)  &    &  $\cdots$               &    &  $\cdots$               &    &  $\cdots$            \\
\hline
Ar$^{4+}$(adopted)       &  1.258$\pm$0.152($-7$)  &1.07&  4.875$\pm$1.115($-8$)  &1.47&  $\cdots$               &    &  $\cdots$               &    &  $\cdots$            \\
\hline
\hline
$[$N\,{\sc I}$]$5198     &  $\cdots$               &    &  $\cdots$               &    &  $\cdots$               &    &  $\cdots$               &    &  4.409$\pm$2.318($-6$)  &    \\
\hline
N$^{0}$(adopted)         &  $\cdots$               &    &  $\cdots$               &    &  $\cdots$               &    &  $\cdots$               &    &  4.409$\pm$2.318($-6$)  &    \\
\hline
\hline
$[$N\,{\sc II}$]$5755    &  1.499$\pm$0.287($-5$)  &    &  6.370$\pm$1.671($-6$)  &    &  2.439$\pm$0.917($-6$)  &    &  1.878$\pm$0.368($-6$)  &    &  1.420$\pm$0.651($-5$)  &    \\
$[$N\,{\sc II}$]$6548    &  1.502$\pm$0.190($-5$)  &    &  6.565$\pm$1.404($-6$)  &    &  2.665$\pm$0.890($-6$)  &    &  1.908$\pm$0.245($-6$)  &    &  1.397$\pm$0.515($-5$)  &    \\
$[$N\,{\sc II}$]$6583    &  1.489$\pm$0.185($-5$)  &    &  6.447$\pm$1.379($-6$)  &    &  2.454$\pm$0.822($-6$)  &    &  1.881$\pm$0.242($-6$)  &    &  1.381$\pm$0.522($-5$)  &    \\
\hline
N$^{+}$(adopted)         &  1.496$\pm$0.391($-5$)  &1.06&  6.470$\pm$2.582($-6$)  &1.38&  2.517$\pm$1.520($-6$)  &1.63&  1.891$\pm$0.504($-6$)  &1.36&  1.396$\pm$0.981($-5$)  &  1.20  \\
\hline
\hline
$[$S\,{\sc II}$]$4069    &  1.676$\pm$0.220($-7$)  &    &  1.660$\pm$0.327($-7$)  &    & 11.030$\pm$3.083($-8$)  &    &  6.943$\pm$0.854($-8$)  &    &  $\cdots$               &    \\
$[$S\,{\sc II}$]$4076    &  $\cdots$               &    &  $\cdots$               &    & 66.702$\pm$19.455($-8$) &    &  $\cdots$               &    &  $\cdots$               &    \\
$[$S\,{\sc II}$]$6716    &  4.107$\pm$0.809($-7$)  &    &  1.804$\pm$0.455($-7$)  &    &  7.198$\pm$2.648($-8$)  &    &  5.674$\pm$0.998($-8$)  &    &  4.394$\pm$2.122($-7$)  &    \\
$[$S\,{\sc II}$]$6730    &  4.605$\pm$0.629($-7$)  &    &  2.012$\pm$0.454($-7$)  &    &  8.048$\pm$2.793($-8$)  &    &  6.386$\pm$0.842($-8$)  &    &  4.947$\pm$2.177($-7$)  &    \\
\hline
S$^{+}$(adopted)         &  4.418$\pm$1.025($-7$)  &1.20&  1.908$\pm$0.643($-7$)  &1.58&  7.600$\pm$3.849($-8$)  &1.92&  6.090$\pm$1.306($-8$)  &1.52&  4.664$\pm$3.040($-7$)  &  1.19  \\
\hline
\hline
$[$S\,{\sc III}$]$6312   &  4.161$\pm$0.575($-6$)  &    &  1.223$\pm$0.291($-6$)  &    &  2.161$\pm$0.764($-6$)  &    &  2.047$\pm$0.298($-6$)  &    &  3.263$\pm$1.118($-6$)  &    \\
\hline
S$^{2+}$(adopted)        &  4.161$\pm$0.575($-6$)  &1.08&  1.223$\pm$0.291($-6$)  &1.47&  2.161$\pm$0.764($-6$)  &1.72&  2.047$\pm$0.298($-6$)  &1.16&  3.263$\pm$1.118($-6$)  &  1.42  \\
\hline
\hline
$[$Cl\,{\sc III}$]$5518  &  4.619$\pm$0.633($-8$)  &    &  $\cdots$               &    &  2.763$\pm$1.015($-8$)  &    &  $\cdots$               &    &  $\cdots$               &    \\
$[$Cl\,{\sc III}$]$5538  &  9.073$\pm$1.108($-8$)  &    &  $\cdots$               &    &  4.145$\pm$1.461($-8$)  &    &  $\cdots$               &    &  $\cdots$               &    \\
\hline
Cl$^{2+}$(adopted)       &  9.073$\pm$1.108($-8$)  &0.90&  $\cdots$               &    &  3.213$\pm$1.780($-8$)  &1.91&  $\cdots$               &    &  $\cdots$               &    \\
\hline
\hline
\end{longtable}
\tablefoot{The adopted notation of the abundance (X$^{i+}$/H$^{+}$), $\alpha\pm\beta(-\gamma)$, means $(a \pm b) \times 10^{-\gamma}$.} 
\end{landscape}
}

\longtab[6]{
\begin{longtable}{lrcrcrcrcrc}
\caption{\label{tab:abun}Total elemental Abundances and Original-to-Revised Ratios, with the Solar Abundances for Comparison.}\\
\hline
\hline
\multicolumn{11}{c}{ICFs follow those by \citet{di2014}}\\
\hline
\multicolumn{1}{c}{Elem.}  &  \multicolumn{1}{c}{M1687}      &  Ratio  &  \multicolumn{1}{c}{M2068}      &  Ratio  &  \multicolumn{1}{c}{M2538}      &  Ratio  &  \multicolumn{1}{c}{M50}      &  Ratio  &  \multicolumn{1}{c}{Z$_{\odot}$}  \\        
\hline																															
He & 10.83$\pm$0.06  &  1.43  & 10.95$\pm$0.10  &  1.22  & 10.96$\pm$0.09  &  1.13  & 11.01$\pm$0.06  &  1.06  & 10.92$\pm$0.02  \\        
N  &  7.32$\pm$0.24  &  2.52  &  8.17$\pm$0.39  &  1.38  &  7.94$\pm$0.34  &  0.99  &  8.07$\pm$0.26  &  1.59  &  7.85$\pm$0.12  \\        
O  &  8.48$\pm$0.12  &  1.10  &  8.58$\pm$0.17  &  1.33  &  8.40$\pm$0.18  &  1.26  &  8.60$\pm$0.13  &  1.05  &  8.73$\pm$0.07  \\  
Ne &  7.87$\pm$0.07  &  1.04  &  8.06$\pm$0.09  &  1.08  &  7.74$\pm$0.10  &  0.95  &  7.91$\pm$0.07  &  1.00  &  8.15$\pm$0.10  \\        
S  &  6.05$\pm$0.22  &  2.88  &  6.68$\pm$0.40  &  1.29  &  6.56$\pm$0.32  &  1.14  &  6.64$\pm$0.25  &  1.06  &  7.15$\pm$0.03  \\        
Cl &  $\cdots$       &$\cdots$&      $\cdots$   &$\cdots$&  $\cdots$       &$\cdots$&  5.05$\pm$0.09  &  1.19  &  5.23$\pm$0.06  \\    
Ar &  5.59$\pm$0.08  &  1.91  &  5.91$\pm$0.10  &  2.09  &  5.48$\pm$0.15  &  2.93  &  5.85$\pm$0.10  &  2.20  &  6.50$\pm$0.10  \\        
\hline                                                              
\multicolumn{1}{c}{Elem.}  &  \multicolumn{1}{c}{M1596}      &  Ratio  &  \multicolumn{1}{c}{M2471}      &  Ratio  &  \multicolumn{1}{c}{M2860}      &  Ratio  &  \multicolumn{1}{c}{M1074}      &  Ratio  &  \multicolumn{1}{c}{M1675}      &  Ratio  \\
\hline                                                              
He & 11.08$\pm$0.03  &  0.98  & 11.00$\pm$0.07  &  1.08  & 10.94$\pm$0.13  &  1.30  & 11.01$\pm$0.05  &  0.97  & 10.98$\pm$0.12  &  1.18  \\
N  &  8.47$\pm$0.20  &  0.87  &  8.11$\pm$0.30  &  0.76  &  7.88$\pm$0.44  &  3.29  &  7.63$\pm$0.21  &  0.88  &  8.42$\pm$0.42  &  2.16  \\
O  &  8.62$\pm$0.10  &  1.07  &  8.54$\pm$0.16  &  1.23  &  8.63$\pm$0.24  &  1.25  &  8.46$\pm$0.11  &  1.12  &  8.64$\pm$0.23  &  1.14  \\
Ne &  7.91$\pm$0.06  &  1.03  &  7.93$\pm$0.09  &  0.95  &  8.01$\pm$0.12  &  0.94  &  7.82$\pm$0.06  &  1.03  &  7.96$\pm$0.12  &  0.94  \\
S  &  7.10$\pm$0.18  &  0.73  &  6.69$\pm$0.27  &  0.77  &  6.49$\pm$0.49  &  3.11  &  6.42$\pm$0.19  &  1.62  &  7.07$\pm$0.47  &  0.96  \\
Cl &  5.30$\pm$0.06  &  0.87  &  $\cdots$       &$\cdots$&  4.85$\pm$0.25  &  2.22  &  $\cdots$       &$\cdots$&  $\cdots$       &$\cdots$\\
Ar &  6.08$\pm$0.07  &  1.95  &  5.73$\pm$0.13  &  2.22  &  5.77$\pm$0.21  &  3.53  &  5.54$\pm$0.07  &  2.10  &  5.83$\pm$0.21  &  2.77  \\
\hline	
\hline
\multicolumn{11}{c}{Using all available X$^{i+}$ abundances with appropriate ICFs}\\
\hline
\multicolumn{1}{c}{Elem.}  &  \multicolumn{1}{c}{M1687}      &  Ratio  &  \multicolumn{1}{c}{M2068}      &  Ratio  &  \multicolumn{1}{c}{M2538}      &  Ratio  &  \multicolumn{1}{c}{M50}      &  Ratio  &  \multicolumn{1}{c}{Z$_{\odot}$}  \\        
\hline																															
He & 10.83$\pm$0.05  &  1.41  & 10.96$\pm$0.09  &  1.20  & 10.96$\pm$0.08  &  1.12  & 11.01$\pm$0.06  &  1.07  & 10.92$\pm$0.02  \\        
N  &  7.36$\pm$0.22  &  2.32  &  8.16$\pm$0.39  &  1.42  &  7.83$\pm$0.35  &  1.27  &  8.00$\pm$0.25  &  1.84  &  7.85$\pm$0.12  \\        
O  &  8.49$\pm$0.12  &  1.08  &  8.60$\pm$0.17  &  1.30  &  8.38$\pm$0.18  &  1.32  &  8.59$\pm$0.14  &  1.07  &  8.73$\pm$0.07  \\        
Ne &  7.87$\pm$0.06  &  1.04  &  8.06$\pm$0.09  &  1.07  &  7.74$\pm$0.10  &  0.95  &  7.91$\pm$0.07  &  1.01  &  8.15$\pm$0.10  \\        
S  &  6.50$\pm$0.09  &  1.03  &  6.64$\pm$0.13  &  1.42  &  6.46$\pm$0.15  &  1.45  &  6.55$\pm$0.10  &  1.29  &  7.15$\pm$0.03  \\        
Cl &  $\cdots$       &$\cdots$&  $\cdots$       &$\cdots$&  $\cdots$       &$\cdots$&  5.05$\pm$0.09  &  1.20  &  5.23$\pm$0.06  \\        
Ar &  5.96$\pm$0.07  &  0.81  &  6.16$\pm$0.09  &  1.19  &  5.86$\pm$0.11  &  1.23  &  6.12$\pm$0.07  &  1.18  &  6.50$\pm$0.10  \\        
\hline                                                              
\multicolumn{1}{c}{Elem.}  &  \multicolumn{1}{c}{M1596}      &  Ratio  &  \multicolumn{1}{c}{M2471}      &  Ratio  &  \multicolumn{1}{c}{M2860}      &  Ratio  &  \multicolumn{1}{c}{M1074}      &  Ratio  &  \multicolumn{1}{c}{M1675}      &  Ratio  \\
\hline                                                              
He & 11.08$\pm$0.03  &  0.98  & 11.00$\pm$0.07  &  1.08  & 10.94$\pm$0.13  &  1.29  & 11.01$\pm$0.05  &  0.97  & 10.98$\pm$0.13  &  1.19  \\
N  &  8.40$\pm$0.19  &  1.02  &  7.99$\pm$0.30  &  1.01  &  7.74$\pm$0.50  &  4.56  &  7.71$\pm$0.19  &  0.74  &  8.33$\pm$0.45  &  2.65  \\
O  &  8.63$\pm$0.10  &  1.05  &  8.53$\pm$0.17  &  1.25  &  8.65$\pm$0.23  &  1.21  &  8.47$\pm$0.10  &  1.11  &  8.65$\pm$0.22  &  1.12  \\
Ne &  7.91$\pm$0.06  &  1.02  &  7.93$\pm$0.09  &  0.96  &  8.01$\pm$0.12  &  0.93  &  7.82$\pm$0.06  &  1.02  &  7.95$\pm$0.11  &  0.95  \\
S  &  6.99$\pm$0.07  &  0.96  &  6.48$\pm$0.12  &  1.24  &  6.65$\pm$0.21  &  2.12  &  6.61$\pm$0.08  &  1.04  &  6.85$\pm$0.16  &  1.57  \\
Cl &  5.31$\pm$0.07  &  0.86  &  $\cdots$       &$\cdots$&  4.83$\pm$0.27  &  2.33  &  $\cdots$       &$\cdots$&  $\cdots$       &$\cdots$\\
Ar &  6.38$\pm$0.05  &  0.99  &  6.07$\pm$0.08  &  1.02  &  6.20$\pm$0.16  &  1.33  &  5.92$\pm$0.06  &  0.87  &  6.18$\pm$0.14  &  1.22  \\
\hline																					
\end{longtable}
}

\end{appendix}

\end{document}